\documentclass[lettersize,journal]{IEEEtran}
\usepackage{amsmath,amsfonts}
\usepackage{algorithmic}
\usepackage{algorithm}
\usepackage{array}
\usepackage[caption=false,font=footnotesize,labelfont=rm,textfont=rm]{subfig}
\usepackage{textcomp}
\usepackage{stfloats}
\usepackage{url}
\usepackage{verbatim}
\usepackage{graphicx}
\usepackage{cite}
\usepackage{color}
\usepackage{amssymb}
\usepackage{multirow}
\usepackage{flushend}
\usepackage{soul}
\usepackage[bb=boondox]{mathalpha}
\usepackage{hyperref} 
\hypersetup{
  colorlinks=true,
  citecolor=blue,
  linkcolor=red,
  urlcolor=green}

\newcommand{\cov}{\boldsymbol{\Theta}}
\newcommand{\bLambda}{\boldsymbol{\Lambda}}

\newcommand{\bgamma}{\boldsymbol{\gamma}}

\newcommand{\rmd}{\mathrm{d}}

\newcommand{\rmA}{\mathrm{A}}

\newcommand{\bh}{\mathbf{h}}

\newcommand{\bb}{\mathbf{b}}

\newcommand{\by}{\mathbf{y}}

\newcommand{\bn}{\mathbf{n}}

\newcommand{\bg}{\mathbf{g}}
\newcommand{\bA}{\mathbf{A}}
\newcommand{\bH}{\mathbf{H}}

\newcommand{\bI}{\mathbf{I}}
\newcommand{\bW}{\mathbf{W}}
\newcommand{\bw}{\mathbf{w}}
\newcommand{\bx}{\mathbf{x}}
\newcommand{\bF}{\mathbf{F}}

\newcommand{\sfn}{\mathsf{n}}

\newcommand{\sfP}{\mathsf{P}}

\newcommand{\sfk}{\mathsf{k}}
\newcommand{\sfR}{\mathsf{R}}

\newcommand{\calO}{\mathcal{O}}
\newcommand{\calR}{\mathcal{R}}
\newcommand{\CN}{\mathcal{CN}}

\newcommand{\bbR}{\mathbb{R}}
\newcommand{\bbE}{\mathbb{E}}

\newcommand{\ub}{\mathsf{UB}}

\newcommand{\calN}{\mathcal{N}}

\newcommand{\ctrans}{\mathsf{H}}
\newcommand{\half}{\frac{1}{2}}

\newcommand{\diag}{\mathrm{diag}}
\newcommand{\Tr}{\mathsf{tr}}

\newcommand{\app}{\mathsf{app}}

\newcommand{\NLoS}{\mathrm{NLoS}}

\newcommand{\eq}{\mathsf{eq}}

\newcommand{\SINR}{\mathrm{SINR}}

\newcommand{\scrU}{\mathfrak{U}}

\newcommand{\noisevar}{\mathcal{\sigma}_{\mathsf{n}}^2}

\newcommand{\BS}{\mathrm{B}}
\newcommand{\UE}{\mathrm{U}}

\newcommand{\fix}{\mathsf{fix}}
\newcommand{\RF}{\mathsf{RF}}
\newcommand{\TC}{\mathsf{DC}}
\newcommand{\BB}{\mathsf{BB}}
\newcommand{\PA}{\mathrm{PA}}

\newcommand{\LNA}{\mathrm{LNA}}
\newcommand{\AD}{\mathrm{AD}}
\newcommand{\DA}{\mathrm{DA}}
\newcommand{\CE}{\mathrm{CE}}
\newcommand{\PC}{\mathrm{P/D}}
\newcommand{\CD}{\mathrm{C/D}}

\newcommand{\DEC}{\mathrm{DEC}}
\newcommand{\UL}{\mathrm{ul}}
\newcommand{\DL}{\mathrm{dl}}
\newcommand{\sub}{\mathrm{sc}}
\newcommand{\ulrat}{\xi^{\UL}}
\newcommand{\dlrat}{\xi^{\DL}}
\newcommand{\PL}{\mathrm{PL}}

\newcommand{\EE}{\mathrm{EE}}
\newcommand{\tot}{\mathrm{sum}}
\newcommand{\syn}{\mathsf{syn}}
\newcommand{\ckt}{\mathsf{circ}}
\newcommand{\RFckt}{\mathrm{RF}\mbox{-}\mathrm{circ}}
\newcommand{\IFckt}{\mathrm{IF}\mbox{-}\mathrm{circ}}

\newcommand{\coef}{\mathcal{I}}

\newcommand{\rL}{\underline{r}}

\newtheorem{proposition}{Proposition}
\newtheorem{remark}{Remark}
\newtheorem{theorem}{Theorem}

\newtheorem{corollary}{Corollary}

\begin{document}

\title{Power Consumption and Energy Efficiency of Mid-Band XL-MIMO: Modeling, Scaling Laws, and Performance Insights}

\author{
Jiachen Tian,~\IEEEmembership{Graduate Student Member,~IEEE,}
Yu Han,~\IEEEmembership{Member,~IEEE,}
Xiao Li,~\IEEEmembership{Member,~IEEE,}\\
Shi Jin,~\IEEEmembership{Fellow,~IEEE,} and
Chao-Kai Wen,~\IEEEmembership{Fellow,~IEEE}
\thanks{J. Tian, Y. Han, X. Li and S. Jin are with the National Mobile Communication Research Laboratory, Southeast University, Nanjing 210096, China (email: \{tianjiachen; hanyu; li\_xiao; jinshi\}@seu.edu.cn).}
\thanks{C.-K. Wen is with the Institute of Communications Engineering, National Sun Yat-sen University, Kaohsiung 80424, Taiwan (e-mail: chaokai.wen@mail.nsysu.edu.tw).}
}

\maketitle

\begin{abstract}
Mid-band extra-large-scale multiple-input multiple-output (XL-MIMO), emerging as a critical enabler for future communication systems, is expected to deliver significantly higher throughput by leveraging the extended bandwidth and enlarged antenna aperture. However, power consumption remains a significant concern due to the enlarged system dimension, underscoring the need for thorough investigations into efficient system design and deployment. To this end, an in-depth study is conducted on mid-band XL-MIMO systems. Specifically, a comprehensive power consumption model is proposed, encompassing the power consumption of major hardware components and signal processing procedures, while capturing the influence of key system parameters. Considering typical near-field propagation characteristics, closed-form approximations of throughput are derived, providing an analytical framework for assessing energy efficiency (EE). Based on the proposed framework, the scaling law of EE with respect to key system configurations is derived, offering valuable insights for system design. Subsequently, extensions and comparisons are conducted among representative multi-antenna technologies, demonstrating the superiority of mid-band XL-MIMO in EE. Extensive numerical results not only verify the tightness of the throughput analysis but also validate the EE evaluations, unveiling the potential of energy-efficient mid-band XL-MIMO systems.
\end{abstract}

\begin{IEEEkeywords}
Mid-band, energy efficiency, near-field communication, power consumption, XL-MIMO.
\end{IEEEkeywords}

\section{Introduction}

\IEEEPARstart{A}{s} wireless communication technologies progress toward the sixth generation (6G), the scarcity of spectrum resources has emerged as a critical bottleneck, constraining the pursuit of higher data rates \cite{6G}. 
Given the limitations of both Sub-6 GHz and millimeter-wave (mmW) frequency bands, increasing attention from academia and industry is shifting toward more promising spectral regions. 
At the World Radiocommunication Conference 2023 (WRC-23), Resolution 245 was issued, recommending the potential global adoption of mid-band spectrum \cite{WRC-23}. 
Therefore, 6G systems are expected to operate within the upper mid-band spectrum, also designated as Frequency Range 3 (FR3) \cite{38820}. 
Beyond spectrum expansion, multi-antenna technologies remain pivotal in enhancing transmission throughput. 
The extra large-scale multiple-input multiple-output (XL-MIMO) system is regarded as an evolution of the massive MIMO system, with enhanced spatial multiplexing capabilities \cite{JYZhang,HQLu}. 
Within the mid-band, a larger number of antennas can be integrated within the same array dimensions compared to Sub-6 GHz systems, which is favorable for implementing XL-MIMO systems \cite{GMIMO}. 
Consequently, the combination of mid-band frequencies and XL-MIMO has emerged as a key trend for future wireless communication systems \cite{WFanCM,Tianmidband,tiansystem}, exploiting the dual advantages in both spectral and spatial domains.
 
Turning to practical deployment, the power consumption of mid-band XL-MIMO systems increases significantly due to the enlarged array size and expanded bandwidth, making energy efficiency (EE) a critical prerequisite.
To achieve efficient deployment of mid-band XL-MIMO systems, a practical power consumption model aligned with the analysis of EE scaling laws with respect to system configurations remains an indispensable research focus. 
Therefore, the evaluation framework serves as a practical foundation for guiding the energy-efficient deployment of mid-band XL-MIMO systems. 

\subsection{Related Works}

Previous research on XL-MIMO systems primarily focused on revealing the remarkable performance gains. 
For instance, the authors in \cite{LuCL} investigated the additional degrees of freedom (DoF) enabled by the near-field propagation characteristics inherent in XL-MIMO systems.
The near-field DoF for a uniform planar array (UPA) configuration was validated in \cite{LDMA}, where the near-field DoF is leveraged for interference suppression.
In \cite{Tianmidband}, the advantages and limitations of near-field propagation were illustrated from the perspectives of different scattering environments.
Beyond theoretical analyses, novel transmission strategies have also been proposed to enhance spectral efficiency (SE) and reduce computational complexity by exploiting unique channel characteristics \cite{XLMIMOJSAC,TianICC}.
In addition to advanced transmission design, EE remains a critical consideration for the practical deployment of mid-band XL-MIMO systems.
Although state-of-the-art works have integrated efficient hardware architectures into XL-MIMO systems, EE has not been thoroughly investigated, hindering efficient system design.
 
Shifting the focus to the mid-band spectrum, existing research primarily concentrates on performance evaluation and propagation characterization.
The authors in \cite{uppermid} analyzed the coverage of the upper mid-band spectrum using ray-tracing tools, illustrating the feasibility and superiority of the mid-band.
For propagation characterization, the mid-band channel characteristics were measured in \cite{U6GELAA}, while key channel metrics are modeled, especially from a distributed perspective.
Comparative measurement campaigns and modeling efforts were presented in \cite{MiaoJSAC}, revealing the potential of the mid-band spectrum in throughput and reliability.
In \cite{JHZhangMag}, the propagation characteristics of the mid-band were comprehensively summarized.
However, EE tailored to the mid-band spectrum remains insufficiently explored, particularly in conjunction with propagation characteristics and wideband features.

 Recalling research on massive MIMO systems in the 5G era, EE emerged as a pivotal metric in performance evaluation.
Power scaling laws under Rayleigh fading and Rician fading were investigated in \cite{EENgo} and \cite{ZhangJSTSP}, respectively, with common findings indicating that the transmit power can be scaled down as the number of antennas grows unboundedly.
However, the hardware power consumption was neglected, resulting in a discrepancy between theoretical and practical EE.
For multiuser communication scenarios, EE-orient optimization with optimal power allocation strategies was studied in \cite{LYou,EEopt}.
Nevertheless, only constant circuit power consumption was considered, neglecting the scaling of system configurations.
Meaningful insights on system design hinge on accurate power consumption characterization \cite{EEComMag}.
Leveraging accurate power consumption models and complete transmission procedures, scaling formulations and advanced algorithms such as antenna selection were considered \cite{EmilTWC,systemJ,AntSel}.
Unfortunately, the aforementioned EE models failed to incorporate key system and channel characteristics of mid-band XL-MIMO systems, thereby calling for throughput and power consumption modeling tailored to this regime.
Moreover, systematic EE comparisons between mid-band XL-MIMO and other representative multi-antenna architectures remain insufficiently explored.

\subsection{Contributions}
Based on the analysis above, EE modeling is indispensable, especially when incorporating the main system and propagation characteristics.
Therefore, the main purpose of this paper is to offer insights into efficient system design for mid-band XL-MIMO systems, grounded in scaling laws derived from a tractable EE analytical framework. 
The contributions can be summarized as follows:
\begin{itemize}
    \item \textit{A comprehensive power consumption model and validation.}
    The mid-band XL-MIMO system exhibits key characteristics including enhanced bandwidth and enlarged array aperture.
    To capture power consumption accurately, the proposed power consumption model incorporates major hardware components and essential signal processing procedures, encompassing practical configurations of typical communication systems.
    Moreover, the proposed model can be regarded as a function of key system configurations, which is also validated by practical data.
    
    \item {\it A tractable analytical framework for EE evaluation.}
    The ergodic throughput for mid-band XL-MIMO systems employing zero-forcing (ZF) precoding and detection is first analyzed, and closed-form upper bounds are subsequently derived.
    By integrating the throughput approximations with the proposed power consumption model, a tractable analytical framework for EE evaluation is established.
    The derived framework not only captures the influences of key system configurations in mid-band XL-MIMO systems but also accommodates various representative multi-antenna technologies, providing a powerful tool for comprehensive EE evaluations and comparisons.
    
    \item {\it In-depth investigations of EE scaling laws with respect to system configurations.}
    Based on the proposed EE analytical framework, scaling laws of EE with respect to key system configurations, mainly including the bandwidth and the number of antennas.
    Beyond theoretical derivations, the validity of these scaling laws is corroborated through extensive numerical evaluations.
    The results demonstrate that achieving an energy-efficient mid-band XL-MIMO system necessitates dedicated design of system parameters, thereby highlighting the practical significance of the proposed EE framework.
\end{itemize} 

\subsection{Organizations and Notations}
The remainder of this paper is organized as follows. 
The system model is introduced in Section \ref{sec:system}. 
The power consumption model and validation is presented in Section \ref{sec:powermodel}.
The proposed EE analytical framework and scaling laws are discussed in Section \ref{sec:EE}.
Comparisons with multi-antenna technologies are given in Section \ref{sec:comparison}.

{\it Notations}--Vectors and matrices are denoted by bold lowercase and uppercase letters, respectively.
The superscripts $(\cdot)^{\top}$ and $(\cdot)^{\mathsf{H}}$ represent the transpose and conjugate transpose, respectively;
$\Tr(\cdot)$ is the trace;
$\odot$ represents the Hadamard product;
$\bbE\{\cdot\}$ stands for the expectation;
$\| \cdot \Vert$ is the $\ell_2$-norm;
$[\bA]_{m,:}$ and $[\bA]_{m,n}$ indicate the elements in the $m$-th row, and the element in the $m$-th row and $n$-th column of matrix $\bA$, respectively; $\diag(\cdot)$ represents a diagonal matrix.
A complex Gaussian distribution is written as $\mathcal{CN}(\boldsymbol{\mu},\boldsymbol{\Sigma})$.
The uniform distribution is denoted as $\scrU(a,b)$.

\section{System Model}
\label{sec:system}

\subsection{System Configuration}

In this paper, both uplink and downlink transmission of a single-cell multiuser mid-band XL-MIMO system are considered, operating over a bandwidth of $B$ Hz. 
The orthogonal frequency division multiplexing (OFDM) is adopted, including $N_{\sub}$ subcarriers with a subcarrier spacing of $\Delta f$.
The block flat-fading channel is assumed, with $B_c$ (in Hz) and $T_c$ (in seconds) representing the coherence bandwidth and coherence time, respectively.
Denoting the duration of an OFDM symbol as $T _{\mathrm{sym}}$, the channels are assumed to be static within a time-frequency coherence block including $S = B_cT_c/(\Delta f T_{\mathrm{sym}})\approx B_cT_c$ resource elements (REs) when ignoring the cyclic prefix.
The base station (BS) is assumed to be located at the center of the cell and is equipped with $N$ half-wavelength spaced antennas,\footnote{Focusing on theoretical analysis, we assume the uniform linear array (ULA) to capture the main characteristics of XL-MIMO without loss of generality. As illustrated in \cite{TanTVT}, the performance under a ULA is comparable to a UPA.} whilst $K$ single antenna users are selected from a large set of users within the coverage area.

\begin{figure}[!t]
    \centering
    \includegraphics[width=0.75\linewidth]{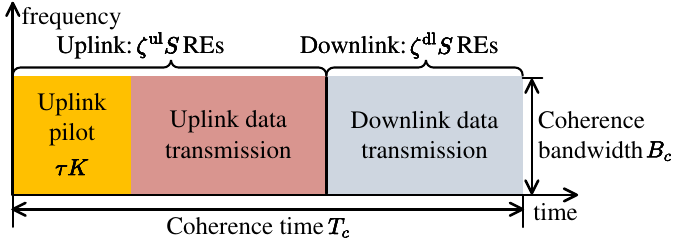}
    \caption{The frame structure of the TDD protocol.}
    \label{fig:frame}
\end{figure}

As shown in Fig. \ref{fig:frame}, the time-division duplex (TDD) protocol is assumed on the basis of perfect synchronization between the BS and the users.
The ratios of uplink and downlink transmission are denoted by $\ulrat$ and $\dlrat$, respectively, satisfying $\ulrat + \dlrat = 1$ when neglecting the duration of TDD switch.
The uplink pilots enable the BS to estimate the channel of the $K$ users. Since the TDD protocol is matched to the coherence blocks, the uplink and downlink channels are considered reciprocal with the assumption of perfect calibration, and the BS can use uplink channel estimates for both reception and downlink transmission. 

\subsection{Channel Model}
The near-field channel model considering large-scale fading is adopted.
The channel of user $k$ is written as
\begin{equation}
    \bh _k = \bgamma _k \odot \tilde{\bh} _k,
\end{equation}
where $\tilde{\bh}_k = \tilde{\cov} _{k} ^{\half} \bg _k$, $\tilde{\cov} _{k}$ is the correlation matrix of small-scale fading, $\bg _{k}$ is a complex Gaussian vector, $\bgamma _k=[\gamma_{k,1},\dots,\gamma _{k,N}] ^{\top}$, with $\gamma _{k,n}$ representing the large-scale fading coefficient from user $k$ to the $n$-th array element, denoted as
\begin{equation}
    \gamma _{k,n} = C_{\PL} \cdot \frac{\lambda}{D_{k,n}},
\end{equation}
where $D_{k,n}$ is the distance between the $k$-th UE and the $n$-th array element, $\lambda$ is the wavelength, and $C _{\PL}$ is a constant value determined by the path-loss model.
The distance term $D _{k,n}$ is further denoted by
\begin{equation}
    D _{k,n} = \sqrt{r _k ^2 + \delta _n ^2 - 2 r _k \delta _n \cos \varphi _k  },
\end{equation}
where $r _k$ and $\varphi _k$ represent the distance and azimuth from the user $k$ to the coordinate origin, respectively, $\delta _n$ is the coordinate of the $n$-th array element.

The small-scale fading correlation matrix can be expressed in a near-field form
\begin{equation}
\begin{aligned}
    \tilde{\cov} _k &= \bbE\{\tilde{\bh} _k \tilde{\bh} _k ^{\ctrans}\} \\
    &=\int_{\tilde{r} _k}\int _{\tilde{\varphi} _k} \bb(\tilde{r} _k, \tilde{\varphi} _k)\bb ^{\ctrans}(\tilde{r} _k,\tilde{\varphi} _k) f(\tilde{r} _k,\tilde{\varphi} _k) \rmd \tilde{r} _k \rmd \tilde{\varphi} _k,
\end{aligned}
\end{equation}
where $f(\tilde{r} _k,\tilde{\varphi} _k)$ captures the channel power distribution of user $k$ in angular and distance domains,\footnote{The power spread in the distance domain is assumed negligible due to the approximate constancy of scatterer ranges within a cluster \cite{Tianmidband, XLMIMOChannel}. Focusing on the angular spread, distributions such as the Von-Mises distribution can be adopted to characterize various propagation regimes of the mid-band through adjustable parameters.} $\bb(r _k,\varphi _k)$ represents the near-field steering vector, specified as
\begin{equation}
    \bb(r _k,\varphi _k) = \left[e^{-\jmath \frac{2\pi}{\lambda} D _{k,1}},\dots,e^{-\jmath \frac{2\pi}{\lambda} D _{k,N}} \right] ^{\top}.
\end{equation}
Therefore, the correlation matrix of the channel is denoted as
\begin{equation}
    \cov _k = \bbE \{\bh _k \bh _k ^{\ctrans}\} = \bbE\{ (\bgamma _k \odot \tilde{\bh} _k)(\bgamma _k \odot \tilde{\bh} _k)^{\ctrans}\} =\boldsymbol{\Gamma} _k \odot \tilde{\cov}_k,
\end{equation}
due to the correlation of the Hadamard produce \cite{MM}, where the matrix $\boldsymbol{\Gamma} _k = \bgamma _k \bgamma _k ^{\ctrans}$ captures the correlation of large-scale fading.

The users are assumed to be uniformly distributed within the cell, characterized by their location parameters, i.e., distances, $r _k$, $k =1, \dots, K$, and azimuths, $\varphi _k$, $k = 1,\dots, K$.
Assuming that the probability density functions (PDFs) of the distance and azimuth are independent, the PDF of distance is modeled as \cite{ParamDist}
\begin{equation}
    f(r_k) = \frac{2 r_k}{r _{\max} ^2 - r _{\min} ^2}, \ r _{\min} \leq r _k \leq r _{\max},
\end{equation}
where $r _{\min}$ and $r _{\max}$ are the minimal distance and cell radius, respectively.
A uniform distribution is adopted for the azimuth, i.e.,
\begin{equation}
    f(\varphi _k) = \frac{1}{\pi}, \ 0 < \varphi _k < \pi,
\end{equation} 
where a semi-circular cell is assumed for simplicity.

\subsection{Channel Estimation}
Before data transmission, a part of the coherence interval is used to acquire channel state information (CSI) at the BS and user sides. 
Channel estimation is usually done through simultaneous transmission of known pilot/training sequences with an overhead of $\tau K$ REs, where the parameter $\tau>0$ can be adjusted to accommodate various pilot occupation patterns.

\subsection{Data Transmission}
Linear signal processing is assumed for uplink data detection and downlink data precoding.
Specifically, ZF detection and precoding are adopted as representative examples. 
For analytical traceability, perfect CSI is assumed at the BS under a high signal-to-noise ratio (SNR) regime, such that the impact of imperfect CSI is negligible. 

In the uplink, the signal received by the BS is given by
\begin{equation}
\by_{\BS} = \sqrt{P ^{\UL}} \bW ^{\ctrans} _{\UL} \bH \bx _{\UE} + \bW _{\UL} ^{\ctrans} \bn _{\BS},
\label{eq:yB}
\end{equation}
where $P ^{\UL}$ is the transmit power of each user under equal power control,\footnote{The transmit power refers to the power per unit bandwidth.} $\bx _{\UE}$ is the user transmit signal with $\bbE \{\bx _{\UE} \bx _{\UE} ^{\ctrans}\} = \bI _{K}$, $\bn _{\BS} \sim \CN(\mathbf{0},\noisevar \bI)$ is the noise, and $\bW _{\UL}$ is the beamforming matrix for ZF detection, given by
\begin{equation}
    \bW _{\UL} = \frac{1}{\sqrt{P ^{\UL}}} \bH (\bH ^{\ctrans} \bH) ^{-1}.
    \label{eq:Wul}
\end{equation}
Thus, the uplink signal-to-interference-plus-noise ratio (SINR) is expressed as
\begin{align}
    \SINR _k ^{\UL} & = \frac{|\bw ^{\ctrans} _{\UL,k} \bh _{k}| ^2} { \sum _{j=1,j\neq k} ^{K} |\bw ^{\ctrans} _{\UL,k} \bh _{j}| ^2 + \| \bw ^{\ctrans} _{\UL,k} \Vert ^2 \noisevar / P ^{\UL} } \notag \\
    & = \frac{P ^{\UL}}{\noisevar [(\bH ^{\ctrans} \bH )^{-1}] _{k,k}},
\end{align}
where $\bw _{\UL,k}$ is the $k$-th column of $\bW _{\UL}$.
Accordingly, the ergodic throughput of the $k$-th user in uplink transmission is
\begin{equation}
    \mathcal{R} _{k} ^{\UL} = \bbE \left\{ \xi ^{\UL} \left( 1 - \frac{\tau K}{S \xi ^{\UL}} \right) B \log_2 ( 1 + \SINR _k ^{\UL}) \right\}.
\end{equation}

In the downlink, the received signal at the user is
\begin{equation}
    \by _{\UE} = \sqrt{P ^{\DL} _{\tot}}\bH ^{\ctrans} \bW _{\DL} \bx _{\BS} + \bn _{\UE},
    \label{eq:yU}
\end{equation}
where $P ^{\DL} _{\tot}$ is the total downlink transmit power, $\bx _{\BS}$ is the transmit signal satisfying $\bbE \{\bx_{\BS} \bx _{\BS} ^{\ctrans}\} = \bI _{K}$, $\bn _{\UE}  \sim  \CN (\mathbf{0},\noisevar \bI)$ is the noise, and $\bW _{\DL}$ is the downlink precoding matrix. For ZF precoding, 
\begin{equation}
    \bW _{\DL} = \bar{\bW} _{\DL} \bLambda =  \bH (\bH ^{\ctrans} \bH) ^{-1} \bLambda,
\end{equation}
where $\bLambda = \diag(\rho _{1},\dots,\rho _{K})$ is the power normalization matrix, 
and $\rho _k \in {\bbR ^{+}}$ is the normalization coefficient for user $k$. To ensure fairness, $\rho _k$ is based on instantaneous channel information \cite{YHanDFT}, given by 
\begin{equation}
    \rho _k = \frac{1}{\sqrt{K} \| \bar{\bw} _{\DL,k} \Vert},
\end{equation}
where $\bar{\bw} _{\DL,k}$ is the $k$-th column of $\bar{\bW} _{\DL}$.
Accordingly, the downlink SINR is 
\begin{equation}
    \SINR _k ^{\DL} \!=\! \frac{P _{\tot} ^{\DL} |\bh ^{\ctrans} _{k} \bw _{\DL, k} |^2}{P _{\tot} ^{\DL} \sum _{j=1,j\neq k} ^{K} |\bh ^{\ctrans} _{k} \bw _{\DL, j} |^2 \! + \! \noisevar} \!=\! \frac{P _{\tot} ^{\DL} \rho _{k} ^2}{\noisevar}.
\end{equation}
Based on the transmission protocol, the ergodic throughput of user $k$ in downlink transmission is
\begin{equation}
    \mathcal{R} _{k} ^{\DL} = \bbE \left\{ \xi ^{\DL} B \log _2(1 + \SINR _k ^{\DL}) \right\}.
\end{equation}

\section{Power Consumption Modeling and validation}
\label{sec:powermodel}
In this section, we aim to provide an accurate characterization of power consumption, which comprises the cumulative contributions of various analog and digital components, radiated power, and signal processing procedures. 
In general, the power consumption model at the BS side and the user side can be expressed as
\begin{equation}
    P _{\bullet} = P_{\bullet,\RF} + P_{\bullet,\TC} + P_{\bullet,\BB} + P_{\bullet,\fix},\  \bullet \in \{\BS,\UE\},
\end{equation}
where $P_{\bullet,\RF}$, $P_{\bullet,\TC}$, $P_{\bullet,\BB}$ and $P_{\bullet,\fix}$ represent the power expenditures of the RF frontend, data converters, baseband signal processing, and a fixed component, respectively.
Note that $P_{\bullet,\fix}$ is a constant term accounting for fixed power consumption required for site cooling, control signaling, and load-independent power of backhaul infrastructure and baseband processors, while the other three components are illustrated as follows. 
In this section, we assume a fully digital architecture is employed at the BS, while the extension to architectures with reduced RF chains is discussed in Section~\ref{sec:mmW}.

\subsection{RF Frontend}
The power consumption of the RF frontend at the BS and an arbitrary user, denoted by $P_{\BS,\RF}$ and $P_{\UE,\RF}$, respectively, can be written as
\begin{subequations}
\begin{align}
    P_{\BS,\RF} &= \xi ^{\DL}P_{\BS,\PA} + N\left(\xi ^{\UL} P_{\LNA} + P_{\syn} + P_{\RF\mbox{-}\ckt}\right), \\
    P _{\UE,\RF} &= \xi ^{\UL}P_{\UE,\PA} + \xi ^{\DL}P_{\LNA} + P_{\syn} + P _{\RF \mbox{-} \ckt},
\end{align}
\end{subequations}
where $P_{\bullet, \PA}$, $\bullet \in \{\BS, \UE\}$ and $P _{\LNA}$ represent the power consumption of the PA and the low-noise amplifier (LNA), respectively. $P _{\syn}$ is the power consumed by a local oscillator (LO), LO amplifier, and mixer, and $P _{\RFckt}$ is the power consumption of other RF circuit components such as PA driver circuits and transmit/receive switches. 
Note that $P _{\syn}$ and $P_{\RFckt}$ are assumed to be bandwidth-independent, and we mainly focus on the power consumption of the PA and LNA. 

Specifically, the power consumption of the PA is determined by the transmit power and the PA efficiency, i.e.,
\begin{subequations}
\begin{align}
    P _{\BS,\PA} &= { B \cdot P _{\tot}^{\DL}}/{\eta _{\PA} ^{\BS}} + NP_{\PA,\mathsf{sta}}, \\
    P _{\UE,\PA} &= {B \cdot P^{\UL}}/{\eta _{\PA} ^{\UE}} + P_{\PA,\mathsf{sta}},
\end{align}
\end{subequations}
where $\eta _{\PA} ^{\UE}$ and $\eta _{\PA} ^{\BS}$ denote the efficiency of the PA at the user and BS sides, respectively, $P _{\PA,\mathrm{bias}}$ represents the static power consumption within the linear region.\footnote{The proposed model aligns well with the linear operating region. In the low input power regime, the equivalent power-added efficiency (PAE) is minimal since the total power consumption is dominated by the static bias power. As the input power increases, the RF output power grows relative to this static overhead, leading to a gradual increase in PAE.}
Although $\eta _{\PA} ^{\UE}$ and $\eta _{\PA} ^{\BS}$ are affected by factors such as amplifier class, it generally holds that $\eta _{\PA} ^{\UE} < \eta _{\PA} ^{\BS}$ due to the more stringent requirements at the user side, including high linearity, support for high-order modulation schemes, and high input power tolerance.
We assume that all users operate within the linear region of their PAs and no explicit power constraints are imposed. 
This assumption is justified by the focus on energy-efficient communication, where users inherently aim to minimize transmit power. 
As a result, the likelihood of exceeding the linear range of the PAs is negligible, rendering the consideration of power constraints unnecessary.
As for the LNA, its power consumption can be modeled as \cite{circuit,LNA}
\begin{equation}
    P _{\LNA} = c_{\LNA}\cdot G_{\LNA} \cdot B
\end{equation}
where $c _{\LNA}$ is the coefficient related to the figure-of-merit (FOM) and noise factor, and $G _{\LNA}$ is the gain of the LNA.

\subsection{Data Converter}
The power consumption of the data converters at the BS and user sides, denoted by $P_{\BS,\TC}$ and $P_{\UE,\TC}$, respectively, can be given as
\begin{subequations}
\begin{align}
    P_{\BS,\TC} &= N\left( \xi^{\UL} P_{\AD} + \xi^{\DL} P_{\DA} + P _{\IFckt}\right), \\
    P _{\UE,\TC} &= \xi^{\DL} P_{\AD} + \xi^{\UL} P_{\DA} + P _{\IFckt},
\end{align}
\end{subequations}
where $P_{\AD}$, $P _{\DA}$ and $P _{\IFckt}$ represent the power consumed by the analog-to-digital converter (ADC), digital-to-analog converter (DAC), and other intermediate frequency (IF) circuit components, respectively. 
Similarly, we assume that the power consumption of IF circuit components is constant, and we focus on modeling the power consumption of the ADC and DAC.

Specifically, the power consumption of an ADC and a DAC handling complex signals with in-phase (I) and quadrature (Q) components can be modeled as \cite{ADC,ADCJSAC,DAC}
\begin{subequations}
\begin{align}
    P_{\AD} &= 2\cdot{c}_{\AD}\cdot 2^{2b _{\AD}} \cdot F_s = 2\epsilon\cdot c_{\AD}\cdot 2^{2b_{\AD}} \cdot B,
    \label{eq:PADC} \\
    P_{\DA} &= 2\cdot c_{\DA}\cdot 2^{2b _{\DA}} \cdot F_s = 2\epsilon \cdot c_{\DA} \cdot 2 ^{2b _{\DA}} \cdot B, 
    \label{eq:PDAC}
\end{align}
\end{subequations}
where $F_s = \epsilon B$ is applied according to the sampling theorem, $\epsilon$ is the oversampling factor, $b_{\AD}$ and $b _{\DA}$ are the number of bits of the ADC and DAC, respectively. The coefficient $c_{\DA}$ is related to ADC architecture and noise factor, while $c_{\DA}$ depends on DAC architecture, noise factor, and FOM. 
Although \eqref{eq:PADC} and \eqref{eq:PDAC} do not fully capture the diversity of practical ADC and DAC architectures, they serve as a fundamental representation of conversion power in relation to key system parameters. 
Thus, incorporating system characteristics such as bandwidth and quantization resolution into the total power consumption model offers a tractable framework for system optimization and performance analysis.

\subsection{Baseband Signal Processing}
\label{sec:SigProces}

When evaluating the power consumption of signal processing, it is common to use the number of computational operations, i.e., floating point operations per second (flops), and computational efficiency \cite{EmilTWC,systemJ,EEZF}. 
Although this model may not precisely reflect the actual power usage, it establishes a link between system configuration and energy consumption through computational complexity. 
Moreover, actual power consumption can be incorporated by adjusting the computational efficiency parameter.

We mainly consider three components: channel estimation, precoding and detection, and channel coding and decoding, which are essential parts of a typical communication system. 
Thus, the power consumption for baseband signal processing at the BS is modeled as
\begin{equation}
    P _{\BS,\BB} = P_{\CE} + P _{\PC} + P_{\CD},
\end{equation}
where $P_{\CE}$, $P _{\PC}$, and $P_{\CD}$ denote the power consumption for channel estimation, precoding and detection, and channel coding and decoding, respectively.

\subsubsection{Channel Estimation}
Channel estimation is performed at the BS based on the uplink pilots. 
For a system with bandwidth $B$ and an observation time of one second, the power consumption can be modeled based on the LS channel estimation as follows:\footnote{The flop counts in \eqref{eq:PCE}, \eqref{eq:PPC}, and \eqref{eq:PDEC} follow the methodology in \cite{matcompute}.}
\begin{equation} 
    P _{\CE} = \frac{B}{B_c} \frac{1}{T_c} \frac{NK (8\tau K -2)}{Q _{\BS}} \approx \frac{B}{S} \frac{8NK^2\tau}{Q _{\BS}},
    \label{eq:PCE}
\end{equation}
where $Q _{\BS}$ denotes the arithmetic capability (in flops/Watt) at the BS. 
Channel estimation at the user side can be neglected due to the single-antenna configuration, which results in only $\mathcal{O}(K)$ complexity.

\subsubsection{Precoding and Combination}
Uplink combining and downlink precoding are also carried out at the BS. 
We consider ZF precoding and combining as an example. The power consumption is modeled as
\begin{multline} 
    P _{\PC} = B\left(1 - \frac{\tau K}{S} \right)(\xi^{\UL}+\xi ^{\DL})\frac{8NK}{Q _{\BS}} \\
    + \frac{B(\xi ^{\UL}+\xi ^{\DL})}{S} \cdot \frac{8K^3/3 + 16NK^2 + 2NK}{Q_{\BS}},
    \label{eq:PPC} 
\end{multline}
where the first term represents the multiplication of the detection matrix $\bW _{\UL}$ with the received signal and the precoding matrix $\bW _{\DL}$ with the transmit signal $\bx _{\UE}$, while the second term accounts for the computations required to generate $\bW _{\UL}$ and $\bW _{\DL}$.
Note that when low-complexity algorithms are employed, the complexity of matrix inversion can be reduced to $\mathcal{O}(K^2)$.

\subsubsection{Coding and Decoding}
The power consumed by channel coding and decoding is mainly determined by transmission throughput \cite{EmilTWC}. 
Specifically, decoding complexity scales with throughput, while encoding complexity is negligible. 
Thus, the power consumption for channel coding and decoding is given by
\begin{equation}
    P_{\CD}=\sum _{k=1} ^{K}\left(R_{k}^{\UL} + R_{k} ^{\DL} \right) Q_{\DEC}/Q_{\BS},
    \label{eq:PDEC}
\end{equation}
where $Q_{\DEC}$ is the number of flops per bit, determined by the employed decoding algorithm.

\subsubsection{Other Signal Processing Procedures} 
In practical mid-band XL-MIMO systems, other signal processing operations also exist.  
Baseband modulation maps raw bits to complex symbols, while demodulation recovers bits from received symbols. 
The associated computational load is quantified as $BKb$ flops, where $b$ is the number of bits per symbol, depending on the modulation scheme.
With OFDM, modulation and demodulation are carried out by inverse fast Fourier transform (IFFT) and FFT, respectively. 
An OFDM symbol is generated by an $N_{\sub}$-point IFFT and recovered by an $N _{\sub}$-point FFT.
Thus, the flops required for OFDM modulation and demodulation in either uplink or downlink are approximately $\calO\left( {5(N + K) B \log_2 N_{\sub} } \right)$ \cite{systemJ}. 
We regard $N _{\sub}$ as a constant,  even as bandwidth scales, since subcarrier spacing typically increases with bandwidth to limit symbol duration, especially in high-speed scenarios and to reduce processing overhead. 

\begin{remark}
    For simplicity, only the essential signal processing procedures are included in Section~\ref{sec:SigProces}, capturing the dominant computational load in a typical mid-band multiuser system. 
    Additional operations can be incorporated by assigning appropriate coefficients. 
\end{remark}
\begin{remark}
    The proposed power consumption model is configurable using practical coefficients and is capable of capturing the impact of key system configurations on EE.
    Furthermore, the proposed model can be extended to hybrid beamforming architectures, due to the modeling at the transceiver channel level, as illustrated in Section \ref{sec:mmW}.
\end{remark}

\begin{figure*}[!t]
    \centering
    \subfloat[Power consumption of RF frontend versus $N$]{\includegraphics[width=0.32\linewidth]{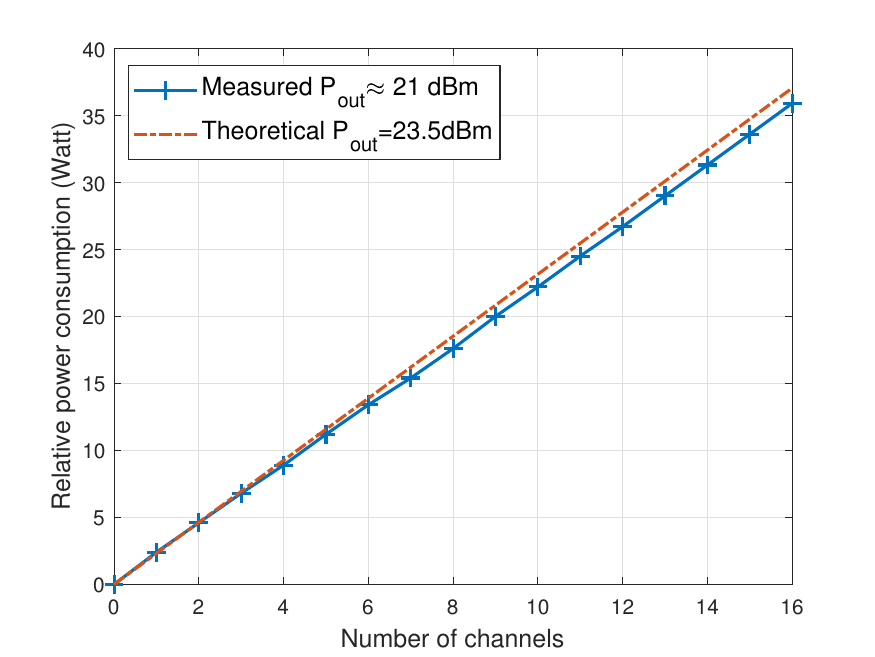}\label{fig:mea_RF}}
    \subfloat[Power consumption of data converter versus $B$]{\includegraphics[width=0.32\linewidth]{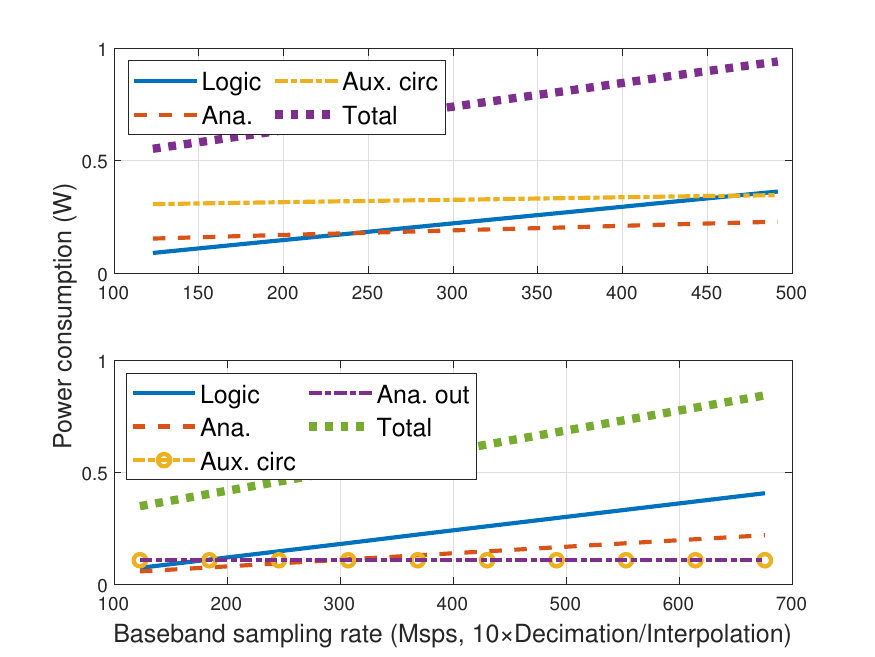}\label{fig:mea_DC}}
    \subfloat[Power consumption of baseband processing]{\includegraphics[width=0.32\linewidth]{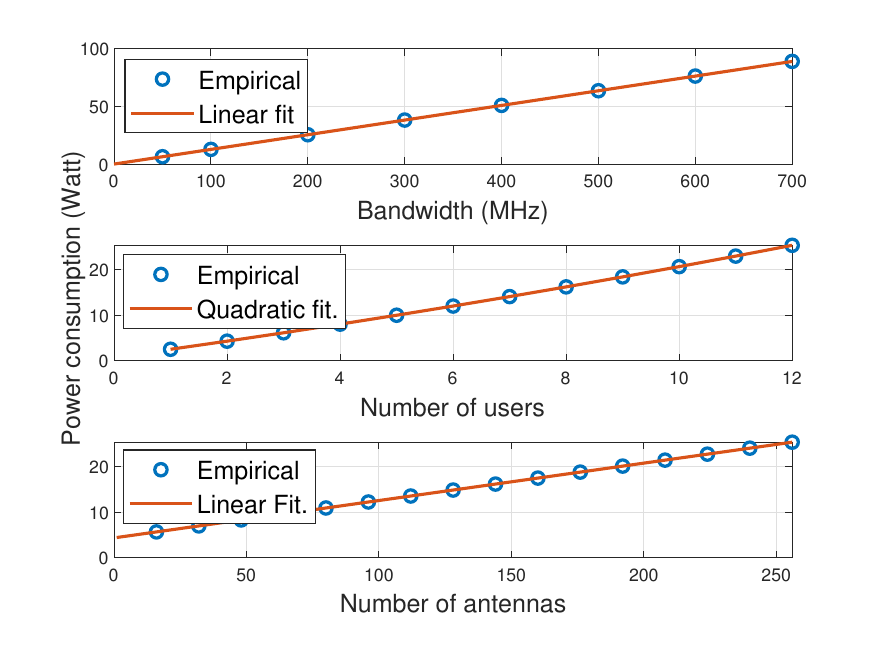}\label{fig:mea_BB}}
 \caption{Validation of the proposed power consumption model.}
 \label{fig:mea}
 \end{figure*}

\subsection{Power Consumption Validation}
The proposed model is intended to capture system-level scaling of power consumption with respect to key parameters.
For instance, the total power consumption scales with $N$ and $B$ linearly.
Accordingly, we carried out validations, with particular emphasis on these scaling relationships.

\subsubsection{RF Frontend}
This part mainly concerns the number of transceiver channels and the bandwidth.
Modeling via power spectral density makes the linear dependence on $B$ inherent, and we mainly examine the power consumption as a function of $N$ based on the RF module in \cite{tiansystem}.
As illustrated in Fig. \ref{fig:mea}\subref{fig:mea_RF}, the measured power scales approximately linearly with $N$, in agreement with the typical values reported in the data sheets, thereby validating the proposed model.

\subsubsection{Data Converter}
The power consumption of data converters depends on system parameters consisting of $N$ and $B$.
Assuming independent data converters, we evaluate the dependence on $B$ via the Xilinx Power Design Manager (PDM), taking the RFSoC ZU47DR as an example.
As shown in Fig. \ref{fig:mea}\subref{fig:mea_DC}, the power consumption of components containing logic, digital processing and auxiliary circuit grows linearly with $B$, implying an affine scaling of the total power consumption.

\subsubsection{Baseband Signal Processing}
Key system parameters in baseband processing include $N$, $K$ and $B$, and we use Xilinx PDM with the Xilinx VU11P FPGAs as the baseband processors.
The total computational load is quantified by the required DSP48s for channel estimation, precoding, and detection \cite{tiansystem}.
As illustrated in Fig. \ref{fig:mea}\subref{fig:mea_BB}, the baseband power consumption is proportional to the total computational load, scaling linearly with $B$ and $N$, and quadratically with $K$, in accordance with the prediction by the proposed model.
 
\section{Proposed Analytical Framework and Scaling Laws of EE}
\label{sec:EE}
Based on the proposed power consumption model, we focus on the analysis of EE in this section.
As the denominator of EE, transmission throughput plays a critical role in the evaluation. 
Therefore, ergodic throughput under typical configurations and propagation conditions of mid-band XL-MIMO systems is first derived, followed by a comprehensive analysis of the associated EE scaling laws.

\subsection{Ergodic Throughput Analysis}
Without loss of generality, we focus on an arbitrary user within the service region and derive an approximation of the ergodic throughput, presented in the following {\it Theorem~\ref{theorem:SE}}. 
Note that the EE analysis should be implemented in a long-term basis, which implies that the throughput should be averaged over all possible user locations.

\begin{theorem}
\label{theorem:SE}
    Based on asymptotic analysis, when ZF detection and precoding are employed at the BS, the upper bounds of the ergodic throughput for user $k$ in uplink and downlink transmission are given by
    \begin{equation}
        \mathcal{R} _{k} ^{\UL,\ub} = \xi ^{\UL} \left( 1 - \frac{\tau K}{S \xi ^{\UL}} \right)B
        {R} _{k} ^{\ub},\ \mathcal{R} _{k} ^{\DL,\ub} = \xi ^{\DL} B {R} _{k} ^{\ub},
        \label{eq:RUB}
    \end{equation}
    where $R^{\ub} _k$ is the upper bound of ergodic SE, expressed as
        \begin{equation}
        {R} _{k} ^{\ub} = \log _2 \left(1 + \frac{P\lambda ^2} {\sigma _n ^2} \left( \chi - (K-1)\frac{I}{\chi} \right) \right),
        \label{eq:RUB}
    \end{equation}
    with $P=P ^{\UL} = P _{\tot} ^{\DL}/K$ representing the average transmit power in both uplink and downlink due to the duality. 
    The terms $\chi$ and $I$ are defined as
    \begin{align}
        \chi &= \sum _{n} \frac{1}{r _{\max} ^2 - r_{\min} ^2} \ln \left( \frac{r_{\max} ^2 - n^2 d_{\rmA} ^2 }{r _{\min} ^2 - n^2 d_{\rmA} ^2} \right),
        \label{eq:Gamma1} \\
        I &= \sum _{n} \left( \frac{1}{r _{\max} ^2 - r_{\min} ^2} \ln \left( \frac{r_{\max} ^2 - n^2 d_{\rmA} ^2 }{r _{\min} ^2 - n^2 d_{\rmA} ^2} \right) \right) ^2.
        \label{eq:I1}
    \end{align}
\end{theorem}

\begin{IEEEproof}
    Refer to Appendix~\ref{appendix:SE}.
\end{IEEEproof}

Although the upper bound characterizes the throughput, the summation over $n$ complicates further analysis with respect to $N$.
Therefore, a more concise upper bound is derived as the approximation in the following {\it Proposition~\ref{coro:SE_UB}}.
\begin{proposition}
\label{coro:SE_UB}
    When ZF detection and precoding are employed, the ergodic throughput for uplink and downlink transmission can be approximated as
    \begin{equation}
        \mathcal{R} _{k} ^{\UL, \app} = \xi ^{\UL}\left( 1 - \frac{\tau K }{S \xi ^{\UL}} \right) B {R} _k ^{\app},\ \mathcal{R} _{k} ^{\DL,\app} = \xi ^{\DL} B {R} _k ^{\app},
    \end{equation}
    where ${R} _k ^{\app}$ is the approximation of the ergodic SE (excluding TDD switching), given by 
    \begin{equation}
        {R} _k ^{\app} = \log _2 \left( 1 + \frac{P \lambda ^2}{\noisevar}\left(\bar{\chi} - {(K-1)}\bar{I}  \right) \right),
        \label{eq:Rapp1}
    \end{equation}
    where $\bar{\chi}$ is denoted in \eqref{eq:barGamma}, shown at the top of page \pageref{eq:barGamma}, and $\bar{I}$ is given as
    \begin{equation}
    \begin{aligned}
        \bar{I} =& \frac{2}{(r_{\max} ^2 - r_{\min}^2)} \Bigg(\ln\left( \frac{r _{\max} + Nd_{\rmA}/2}{r _{\min} + Nd_{\rmA}/2} \right) \\
        & + \frac{N d_{\rmA}}{2} \left(\frac{1}{r_{\max} \!+\! {N  d_{\rmA}}/{2}} \!-\! \frac{1}{r_{\min} \!+\! {N  d_{\rmA}}/{2}} \right)\Bigg).
        \label{eq:I_bar}
    \end{aligned}
    \end{equation}
    
    \newcounter{TempEqCnt}
    \setcounter{TempEqCnt}{\value{equation}}
    \begin{figure*}[!t]
    \normalsize
    \begin{equation}
    \bar{\chi}  =  \frac{1}{r _{\max} ^2 - r_{\min} ^2}\Bigg ( N\ln\left( \frac{4r_{\max}^2 / d_{\rmA} ^2 - N^2}{4r_{\min}^2/d_{\rmA}^2 - N^2} \right) +  \frac{2r_{\max}}{d_{\rmA}} \ln \left( \frac{2r_{\max} / d_{\rmA} + N}{2r_{\max} / d_{\rmA} - N} \right) - \frac{2r_{\min}}{d_{\rmA}} \ln \left( \frac{2r_{\min} / d_{\rmA} + N}{2r_{\min} / d_{\rmA} - N} \right) \Bigg).
    \label{eq:barGamma}
    \end{equation}
    \hrulefill
    \end{figure*}
\end{proposition}

\begin{IEEEproof}
    Refer to Appendix~\ref{appendix:SE_UB}.
\end{IEEEproof}

In this paper, we evaluate EE from a long-term perspective and focus on network-level average throughput. The derived upper bounds remain applicable under various small-scale fading models. In view of the wide mid-band spectrum, the proposed bounds capture both rich-scattering and highly directional propagation characteristics, making them suitable for mid-band scenarios. Moreover, the proposed bounds offer tractable throughput approximations that explicitly account for the main system parameters.

\begin{figure}[!t]
    \centering
    \includegraphics[scale=0.45]{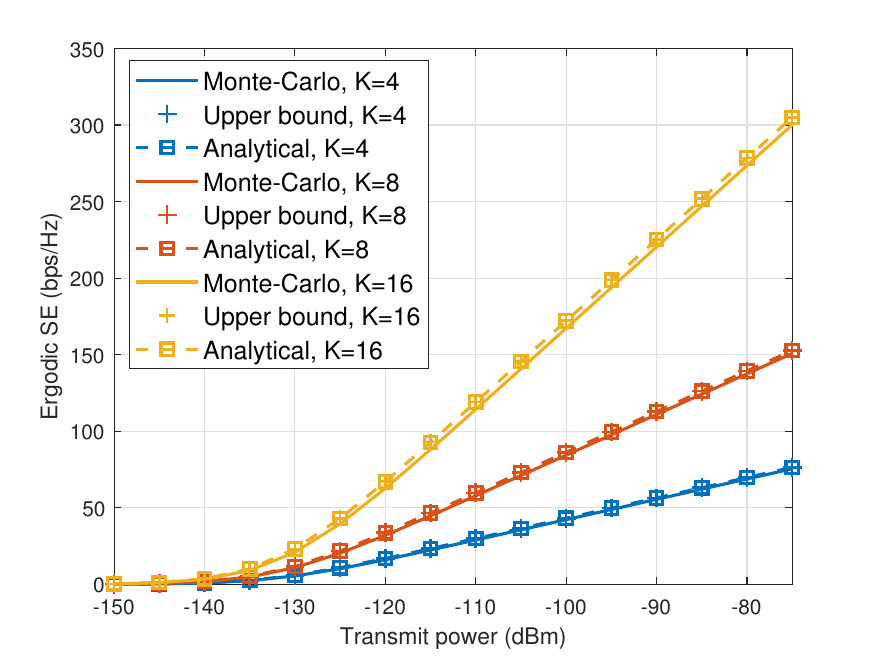}
    \caption{The upper bound, approximation and Monte-Carlo result of SE for different numbers of users.}
    \label{fig:SE_midband}
\end{figure}

The tightness of the derived expressions is shown in Fig.~\ref{fig:SE_midband}, which compares the upper bound in \eqref{eq:RUB}, the approximation in \eqref{eq:Rapp1}, and Monte Carlo simulation results under various user numbers for the mid-band XL-MIMO system.
The simulation parameters follow those listed in Table~\ref{tab:param1}, and the channel correlation matrices are generated based on the approximation method in \cite{TianICC}.
It is observed that both the upper bound and the approximation closely match the Monte Carlo results, confirming the tightness and validity of the proposed analytical framework for ergodic SE.
 
\subsection{Scaling Laws of EE}

\begin{table}[!t]
\renewcommand{\arraystretch}{1.2}
	\caption{Coefficients in the proposed power consumption model \label{tab:coef}}
	\centering
	\begin{tabular}{m{1cm}<{\centering} m{7cm}<{}}
	\hline
	Coefficient & Expression \\
	\hline
        $\coef _{\sfn,0}=$ & $c_{\LNA} \xi^{\UL} G_{\LNA} B + P_{\syn} + P_{\RFckt} + P_{\IFckt} + 2\epsilon B \left(\xi^{\UL} c_{\AD} 2^{2b_{\AD}} +\xi^{\DL} c_{\DA} 2^{2b_{\DA}}\right)+{P} _{\PA, \mathsf{sta}}$\\
        $\coef _{\sfn,1}=$ & $B \cdot \frac{8 (\xi^{\UL} + \xi^{\DL})}{Q_{\BS}}$\\
        $\coef _{\sfn,2}=$ & $\frac{B}{S Q_{\BS}} \cdot \left(16 (\xi^{\UL} + \xi^{\DL}) + 8 \tau (\xi^{\UL} + \xi^{\DL}) - 8 \tau\right)$\\
        $\coef _{\sfk,1}=$ & $c_{\LNA} \xi^{\DL} G_{\LNA} B \!+\! P_{\syn} \!+\! P_{\RFckt} \!+\! P_{\IFckt} \!+\! 2\epsilon B \left(\xi^{\DL} c_{\AD} +  2^{2b_{\AD}} + \xi^{\UL} c_{\DA} 2^{2b_{\DA}}\right) \!+\! P_{\UE,\fix}+{P} _{\PA, \mathsf{sta}}$\\
        $\coef _{\sfk,3}=$ & $\frac{8 B \left( \xi^{\UL} + \xi^{\DL} \right)}{3 S Q_{\BS}}$\\
        $\coef _{\fix} = $ & $P _{\BS, \fix}$ \\
        $\coef _{\sfP} = $ & $B\cdot\left( \frac{\xi ^{\UL}}{\eta _{\PA} ^{\UE}} + \frac{\xi ^{\DL}}{\eta _{\PA} ^{\BS}} \right)$ \\
        $\coef _{\sfR} = $ & $Q _{\DEC}/Q_{\BS}$ \\
        \hline
	\end{tabular}
\end{table}

Recalling the power consumption model in Section~\ref{sec:powermodel}, the total power consumption of a multiuser mid-band XL-MIMO system is given by
\begin{equation}
    P_{\tot} = P_{\BS} + KP _{\UE}.
\end{equation}
which can be expressed explicitly as in \eqref{eq:Ptot}, shown at the top of page \pageref{eq:Ptot}, where the coefficients $\coef _{\sfP}$, $\{\coef _{\sfn,i}\} _{i=0} ^{2}$, $\{\coef _{\sfk,i}\} _{i=0,1,3}$, $\coef _{\sfR}$ and $\coef _{\fix}$ are listed in Table \ref{tab:coef}.  

\setcounter{TempEqCnt}{\value{equation}}
\begin{figure*}[!t]
\normalsize
\begin{equation}
P_{\tot} = \coef _{\sfP}KP + N \sum _{i=0,1,2} \coef _{\sfn,i} K ^{i} + \sum _{i=1,3} \coef_{\sfk,i} K ^{i} + \coef _{\sfR} BK\left(1 - \frac{\tau K}{S} \right) {R} _k ^{\app} +\coef_{\fix}.
    \label{eq:Ptot}
\end{equation}
\hrulefill
\end{figure*}

Combining this with the tractable expression for ergodic throughput, the EE can be approximated as
\begin{equation}
    \EE \approx BK\left(1 - \frac{\tau K}{S} \right) \frac{ {R} _{k} ^{\app}} {P _{\tot}}.
    \label{eq:EE}
\end{equation}
Equation~\eqref{eq:EE} establishes a tractable analytical framework for EE evaluation and analysis. 
Based on this framework, the scaling of EE with respect to key system configurations in mid-band XL-MIMO systems can be derived.
The increased number of antennas and enhanced bandwidth are characteristics distinguished from other systems, which are the focus of this paper.
The derived scaling formulas are further validated through numerical results presented in the following sections.

\setcounter{TempEqCnt}{\value{equation}}
\begin{figure*}[!t]
\normalsize
\begin{equation}
\lim_{B\rightarrow \infty}\EE = \frac{K\left(1 - \frac{\tau K}{S} \right) R_{k}^{\app}}  
{\bar{\coef} _{\sfP}KP + N \sum _{i=0,1,2} \bar{\coef} _{\sfn,i} K ^{i} + \sum _{i=1,3} \bar{\coef} _{\sfk,i} K ^{i} + \coef _{\sfR} K\left(1 - \frac{\tau K}{S} \right) {R} _k ^{\app} }.
\label{eq:EE_BW_lim}
\end{equation}
\hrulefill
\end{figure*}

\begin{table*}[!t]
	\caption{Parameter Setting for EE Evaluation \label{tab:param1}}
	\centering
	\begin{tabular}{l|l||l|l}
	\hline
	Parameter & Value & Parameter & Value \\
	\hline
        Number of antennas $N$ & 512 & Carrier wavelength $f_c$ & 7.5 GHz\\
        Cell radius (maximal distance) $r_{\max}$ & 150 m & Minimum distance $r _{\min}$ & 70 m \\
        Bandwidth $B$ & 400 MHz & Computational efficiency at the BS $Q _{\BS}$ & 30 Gflops/W\\
        REs in a coherence block $S$ & 1000 & Normalized coefficient of pathloss $C _{\PL}$ & 1 \\
        Ratio of CSI acquisition $\tau$ & 1 & PA static power consumption $P _{\PA,\mathsf{sta}}$ & 300 mW\\
        Fraction of uplink transmission $\xi ^{\UL}$ & 0.4 & Fraction of uplink transmission $\xi ^{\DL}$ & 0.6 \\
        Fixed power consumption at the BS $P _{\BS,\fix}$ & 15 W & Fixed power consumption at the UE $P _{\UE,\fix}$ & 2 W \\
        Coefficient of ADC $c_{\AD}$ & $1.97\times 10^{-19}$ Joule/step& Coefficient of DAC $c_{\DA}$ & $1.66\times10^{-19}$ Joule/step \\
        Number of bits in ADC $b _{\AD}$ & 14 & Number of bits in DAC $b _{\DA}$ & 14 \\ 
        Coefficient of the LNA $c _{\LNA}$ & $1.67 \times 10 ^{-11}$ W/Hz & Typical gain of the LNA $G$ & 20 dB \\
        Efficiency of the PA at the BS side $\eta _{\PA} ^{\BS}$ & 30\% & Efficiency of the PA at the user side $\eta _{\PA} ^{\UE}$ & 15\% \\
        IF driver circuits $P _{\IFckt}$ & 300 mW & RF driver circuits $P _{\RFckt}$ & 500 mW\\
        Power consumed by LO and mixer $P _{\syn}$ & 50 mW & Channel decoding flops $Q _{\DEC}$ & 100 flops/bit\\
        \hline
	\end{tabular}
\end{table*}

\begin{figure}[!t]
    \centering
    \includegraphics[scale=0.45]{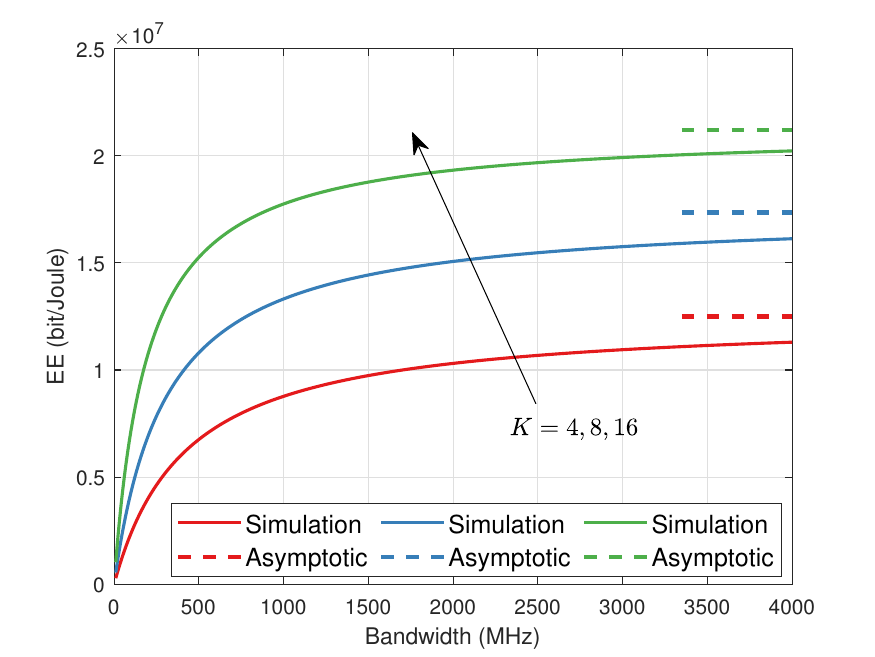}
    \caption{EE versus the bandwidth.}
    \label{fig:EE_vs_BW}
\end{figure}

\begin{table}[!t]
\renewcommand{\arraystretch}{1.2}
	\caption{Coefficients in \eqref{eq:EE_BW_lim} \label{tab:coef_BW}}
	\centering
	\begin{tabular}{m{1.3cm}<{\centering} m{6.2cm}<{}}
	\hline
	Coefficient & Expression \\
	\hline
        $\bar{\coef} _{\sfn,0}=$ & $c_{\LNA} \xi^{\UL} G_{\LNA} + 2\epsilon \left(\xi^{\UL} c_{\AD} 2^{2b_{\AD}} +\xi^{\DL} c_{\DA} 2^{2b_{\DA}}\right)$\\
        $\bar{\coef} _{\sfn,1}=$ & $\frac{8 (\xi^{\UL} + \xi^{\DL})}{Q_{\BS}}$\\
        $\coef _{\sfn,2}=$ & $\frac{1}{S Q_{\BS}} \cdot \left(16 (\xi^{\UL} + \xi^{\DL}) + 8 \tau (\xi^{\UL} + \xi^{\DL}) - 8 \tau\right)$\\
        $\bar{\coef} _{\sfk,1}=$ & $c_{\LNA} \xi^{\DL} G_{\LNA} + 2\epsilon \left(\xi^{\DL} c_{\AD} 2^{2b_{\AD}} + \xi^{\UL} c_{\DA} 2^{2b_{\DA}}\right)$\\
        $\bar{\coef} _{\sfk,3}=$ & $\frac{8 \left( \xi^{\UL} + \xi^{\DL} \right)}{3 S Q_{\BS}}$\\
        $\bar{\coef} _{\sfP} = $ & $\left( \frac{\xi ^{\UL}}{\eta _{\PA} ^{\UE}} + \frac{\xi ^{\DL}}{\eta _{\PA} ^{\BS}} \right)$ \\
        \hline
	\end{tabular}
\end{table}

\subsubsection{Bandwidth}
In general, the power consumption of both circuit components and baseband signal processing is closely related to the system bandwidth $B$.
However, from the expression of EE in \eqref{eq:EE}, it can be observed that although the total power consumption increases significantly with the bandwidth, the EE may remain at an acceptable level due to the corresponding improvement in throughput, as formalized in \textit{Corollary}~\ref{Coro:Bandwidth}.
\begin{corollary}
    When the other system configurations are fixed, the EE increases with increasing $B$ and converges to a constant, given in \eqref{eq:EE_BW_lim}, at the top of page \pageref{eq:EE_BW_lim}, where the parameters are listed in Table \ref{tab:coef_BW}.
    \label{Coro:Bandwidth}
\end{corollary}
\begin{IEEEproof}
    The throughput grows $\propto B$ and the power consumption is affine in $B$, and the EE rises with $B$ and eventually saturates to a constant value.
\end{IEEEproof}

As shown in Fig.~\ref{fig:EE_vs_BW}, the relationship between EE and system bandwidth under different numbers of users is illustrated.
It can be observed that the EE increases with the bandwidth and gradually converges to asymptotic values, thereby validating \textit{Corollary}~\ref{Coro:Bandwidth}.
Summarizing the numerical results and \textit{Corollary}~\ref{Coro:Bandwidth}, we derive the following insights on system design.

\textit{Insights:} Under the proposed EE model, increasing the bandwidth enhances both throughput and EE.
According to the asymptotic expression, EE is primarily determined by hardware performance.
Therefore, given the predefined bandwidth, EE can be improved by deploying advanced hardware components, such as more efficient computational resources and data converters.
Moreover, when the bandwidth is 400~MHz, a commonly expected value for mid-band XL-MIMO systems in industrial blueprints such as \cite{samsung}, the EE approaches a relatively high value.

\begin{figure}
    \centering
    \includegraphics[scale=0.45]{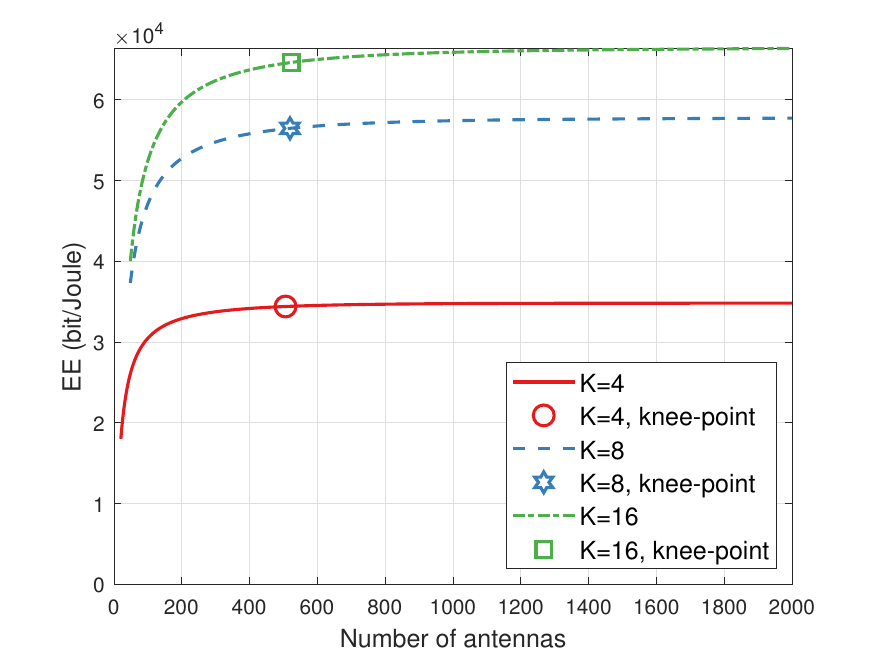}
    \caption{EE versus the number of antennas.}
    \label{fig:EE_vs_numAnt}
\end{figure}

\subsubsection{Number of Antennas}
As another significant factor in mid-band XL-MIMO systems, increasing the number of antennas serves as another powerful enabler for the throughput.
To investigate the energy cost of introducing more antennas, we detail the scaling of EE w.r.t. the number of antennas on the foundation of the throughput scaling in {\it Proposition}\,\ref{prop:antenna}.

\begin{proposition}
\label{prop:antenna}
    When the ZF precoding and detection are employed, the scaling laws of the array gain $\bar{\chi}$ and throughput are as follows.
    For $N\ll 2r _{\min}/d_\rmA$, $\bar{\chi} $ increase approximately linearly with $N$ and can be expressed as
    \begin{equation}
        \bar{\chi} = \frac{2\ln \left( {r_{\max}}/{ r _{\min}} \right)}{r_{\max}^2 - r_{\min}^2} N + \mathcal{O}(N).
    \end{equation}
    When $N$ approaches $2r _{\min}/d_\rmA$, $\chi$ saturates to the finite limit \eqref{eq:chi_limit}, shown at the top of page \pageref{eq:chi_limit}.

    \setcounter{TempEqCnt}{\value{equation}}
    \begin{figure*}[!t]
    \begin{equation}
    \lim_{N \to \left(\frac{2r_{\min}}{ d_{\rmA} }\right)^{-}} \! \bar{\chi} = \frac{1}{r_{\max}^2 \!-\! r_{\min}^2} \Bigg( \frac{2 r_{\max}}{d_{\rmA}}\ln\left( \frac{r_{\max} \!+\! r_{\min}}{r_{\max} \!-\! r_{\min}} \right) + \frac{2 r_{\min}}{d_{\rmA}}\ln\left( \frac{r ^2 _{\max} - r ^2 _{\min}}{4r^2_{\min}} \right) \Bigg).
    \label{eq:chi_limit}
    \end{equation}
    \hrulefill
    \end{figure*}

    Regarding the scaling of throughput, we focus on the array gain and neglect the interference term for clarity.
    In the low transmit power regime with $N\ll 2r _{\min}/d_\rmA$, the throughput grows approximately linearly with $N$ and satisfies
    \begin{equation}
        \calR \sim  \frac{2BK\left(1-\frac{\tau K}{S}\right)P \lambda^2 }{\noisevar\ln2 (r_{\max}^2 - r_{\min} ^2) } \ln\left( \frac{r_{\max}}{r _{\min}} \right)N.
        \label{eq:R_LTx}
    \end{equation}
    Whilst for high transmit power, when $N \ll 2r _{\min}/d_\rmA$, the throughput logarithmically with $N$, i.e., 
    \begin{equation}
        \calR \!\sim\! BK\left(1 \!-\! \frac{\tau K}{S}\right)\log_2\left(1 \!+\! \frac{2PN \lambda^2 \ln \left( \frac{r_{\max}}{ r_{\min} } \right) }{\noisevar(r_{\max}^2 - r_{\min}^2)} \right).
        \label{eq:R_HTx}
    \end{equation}
\end{proposition}

\begin{IEEEproof}
    Refer to Appendix \ref{appendix:antenna}.
\end{IEEEproof}

Based on the analysis of throughput above, we then delve into the scaling of EE.

\begin{corollary}
\label{Coro:N}
    Regarding the total power consumption $P_{\tot}$ scales unboundedly with $N$, the EE decreases with the number of array elements at high transmit power and satisfy
    \begin{equation}
        \lim _{N\rightarrow +\infty} \EE = 0.
    \end{equation}
    Under low transmit power, the EE initially increases, saturates near a plateau, and subsequently decreases, with an approximate knee-point given as
    \begin{equation}
        N^{\mathsf{kp}} = \frac{\eta_{\sfn}}{1-\eta_{\sfn}} \cdot \frac{\coef _{\sfP}KP + \sum _{i=1,3} \coef_{\sfk,i} K ^{i}}{\sum _{i=0,1,2} \coef _{\sfn,i} K ^{i}},
        \label{eq:x_kp}
    \end{equation}
    where $\eta_{\sfn}$ denotes the percentage of the plateau value.
\end{corollary}

\begin{IEEEproof}
Refer to Appendix~\ref{appendix:N}.
\end{IEEEproof}

\setcounter{TempEqCnt}{\value{equation}}
\begin{figure*}[!t]
\begin{equation}
\SINR _{\ell k} = \frac{P_{\ell k}}{\noisevar[(\bH _{\ell \ell} ^{\ctrans} \bH _{\ell \ell}) ^{-1}]_{k,k} + \sum_{i\neq \ell} \sum_{u=1}^{K}P _{iu} [\bW ^{\ctrans} _{\UL,\ell} (\bh_{\ell iu} \bh _{\ell iu} ^{\ctrans}) \bW _{\UL,\ell}] _{k,k}} 
\overset{\mathrm{(a)}}{\approx} \frac{P _{\ell k}}{[(\bH _{\ell \ell} ^{\ctrans} \bH _{\ell \ell}) ^{-1}]_{k,k}(\noisevar + P _{iu}  \sum _{i\neq \ell}\sum_{u=1}^{K}\gamma ^2 _{\ell i u})},
\label{eq:SINR_mc}
\end{equation}
\hrulefill
\end{figure*}

As verified in Fig.~\ref{fig:EE_vs_numAnt}, at low transmit power $P = -150\, \mathrm{dBm}$, the EE first rises and then falls with the number of antennas. The knee point, defined as the antenna count at which the EE reaches $\eta_{\sfn}=95\%$ of its plateau, marks a sufficient array size.
Increasing antennas can initially boost EE, and the trend eventually saturates.
The subsequent decrease is primarily driven by the power consumption associated with each antenna (transceiver channel).
From another perspective, due to the near-field effects, particularly the increased angular span \cite{LuICC}, which degrades inter-user interference cancellation. Unlike far-field conditions, where adding antennas typically enhances ergodic throughput, this reveals a fundamental shift in scaling behavior. 
Based on the analysis above, we derive the following insights.

{\it Insight:} When the transmit power is limited, increasing the antennas benefits both throughput and EE, whilst the required number of antennas exceeds that of conventional massive MIMO deployments, thereby underscoring the necessity of extra large-scale arrays. 
For high transmit power regimes, the number of antennas should be determined based on the requirements of other indicators, such as ecoverage and throughput, as adding more antennas imposes an EE penalty.

\subsection{Extension to the Multi-cell Scenario}
\label{sec:discuss}

The analysis so far has been based on a single-cell scenario.
We now extend the framework to a multi-cell setup.
Assuming a total of $L$ cells, the uplink signal received at the BS of cell $\ell$, with ZF detection, is expressed as
\begin{equation}
    \by _{\BS,\ell} = \bx _{\UE,\ell} + \sum _{i=1,i\neq \ell} ^{L} \bW ^{\ctrans} _{\UL,\ell} \bH _{\ell i} \bx _{\UE,i} + \bW ^{\ctrans} _{\UL,\ell} \bn _{\BS,\ell},
\end{equation}
where $\bW _{\UL,\ell} = \bH _{\ell \ell} (\bH _{\ell \ell} ^{\ctrans} \bH _{\ell \ell}) ^{-1} $, $\bH _{\ell i}$ denotes the channel from users in cell $i$ to the BS in cell $\ell$, and $\bx_{\UE,i}$ is the transmit signal from users in cell $i$.
The corresponding SINR for user $k$ is denoted as \eqref{eq:SINR_mc}, where $\gamma_{\ell i u}$ is the pathloss from user $k$ in cell $i$ to cell $\ell$, (a) is due to the far-field assumption for the inter-cell channels, and the inter-cell channel is modeled as $\bh_{\ell i u} \sim \mathcal{CN}(\mathbf{0},\gamma_{\ell i u}^2 \bI)$. 
We adopt the modeling because we mainly focus on the influence of other cells on the focused cell, captureing the inter-cell interference from distant users under rich-scattering propagation without a line-of-sight component.
It can be derived that the EE scales with the bandwidth following a similar relationship to the single-cell scenario, although the asymptotic value is reduced due to the lower SE. 
Similarly, with respect to the number of antennas, the EE scaling trend remains consistent with that of the single-cell case, while the EE at the knee point is reduced owing to inter-cell interference.

\section{Comparisons With Representative Multi-antenna Technologies}
\label{sec:comparison}

As a promising enabler among multi-antenna technologies, the superiority of midband XL-MIMO systems in throughput is reflected from the enhanced bandwidth in midband spectrum and the enlarged antenna array.
However, how these features influence the EE should be further demonstrated through comparisons with other representative architectures.
From another perspective, heterogeneous networking that integrate various multi-antenna technologies are considered promising for future wireless systems.
Therefore, two representative system configurations are considered in this section, differing from the mid-band XL-MIMO system in terms of array size, propagation characteristics, and system architecture.\footnote{In the following sections, variables denoted with an overline and an underline refer to the two comparative systems, defined in a manner consistent with the mid-band XL-MIMO system.}

\subsection{MIMO System on Sub-6 GHz Band}
Since the 4G era, MIMO systems employing multiple antennas have become a fundamental technology, often operating in the Sub-6 GHz frequency band.
Consider a MIMO system with bandwidth $\underline{B}$, where the BS is equipped with an antenna array of $\underline{N}$ elements and serves $\underline{K}$ single-antenna users.
For the propagation model under the sub-6GHz band, the channel vector between the BS and an arbitrary user $k$ is given by
\begin{equation}
    \underline{\bh} _k = \underline{\gamma} _k \cdot \underline{\bg} _{k},
\end{equation}
where $\underline{\gamma} _{k} = {\underline{\lambda}} / {\underline{r} _k}$ denotes the large-scale fading coefficient, and $\underline{\bg} _k$ is a channel vector with i.i.d. complex Gaussian distribution $\CN(0,1)$.
Based on this model and using a signal model consistent with \eqref{eq:yB} and \eqref{eq:yU}, a closed-form approximation of the ergodic throughput is derived as follows.
\begin{proposition}
\label{prop:Rsub6}
    Given the configurations of a MIMO system, the ergodic SE of an arbitrary user $k$ is lower-bounded by
    \begin{equation}
        \begin{aligned}
        \underline{R} _k =&\frac{1}{(\rL _{\max} ^2 \!-\! \rL _{\min} ^2)\ln 2} \left( \underline{C} \ln \left( \frac{ \rL _{\max} ^2 \!+\! \underline{C} }{\rL _{\min} ^2 \!+\! \underline{C}} \right) \right.\\
        &\left.+ \rL _{\max} ^2\ln \left( \frac{ \rL _{\max} ^2 \!+\! \underline{C} }{\rL _{\max} ^2 \!+\! \underline{C}} \right) \!-\! \rL _{\min} ^2 \left ( \frac{  \rL _{\min} ^2 \!+\! \underline{C}}{\rL _{\min} ^2 \!+\! \underline{C}} \right) \right),
        \end{aligned}
        \label{eq:Rsub6}
    \end{equation}
    where $\underline{C}=\underline{P}\underline{\lambda}^2 (\underline{N} - \underline{K}) / \noisevar$.
\end{proposition}

\begin{IEEEproof}
    Refer to Appendix~\ref{appendix:Rsub6}.
\end{IEEEproof}

\begin{figure}[!t]
    \centering
    \includegraphics[scale=0.45]{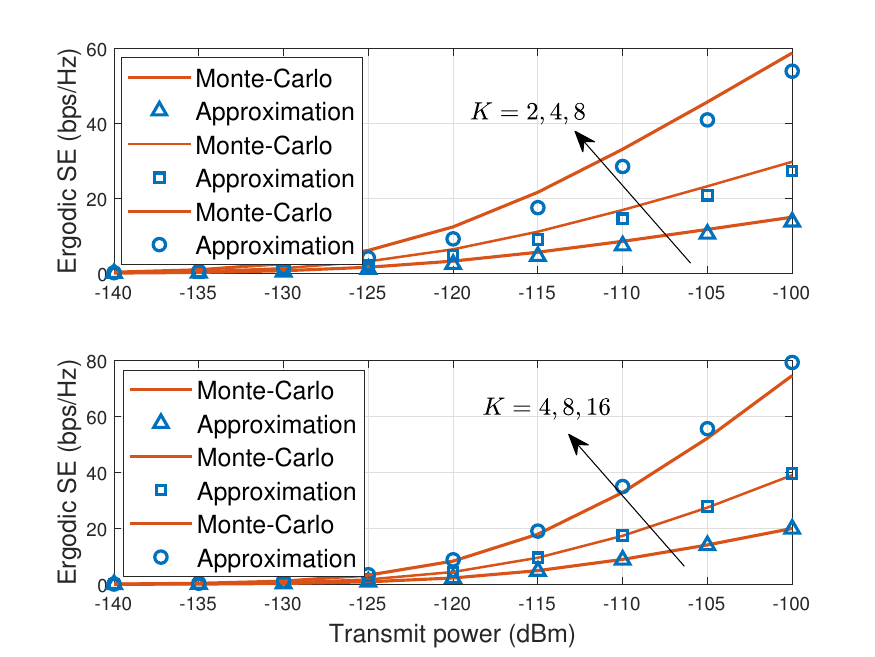}    \caption{The approximation and Monte-Carlo result of SE for different representative multi-antenna technologies. Top: Setup 1. Buttom: Setup 3 with $N _{\RF} = \bar{K}$.}
    \label{fig:SE_submmW}
\end{figure}

The power consumption model for Sub-6 GHz MIMO systems follows a similar structure to that of the mid-band XL-MIMO system.
The total power consumption can be expressed as in \eqref{eq:Ptot_sub6}, shown at the top of page~\pageref{eq:Ptot_sub6}, where the coefficients $\underline{\coef} _{\sfP}$, $\{\underline{\coef} _{\sfn,i}\} _{i=0} ^{2}$, $\{\underline{\coef} _{\sfk,i}\} _{i=0,1,3}$, $\underline{\coef} _{\sfR}$ and $\underline{\coef} _{\fix}$ are defined analogously to those in Table~\ref{tab:coef}, but based on parameters appropriate for the Sub-6 GHz band. 
Accordingly, the EE can be approximated as
\begin{equation}
    \underline{\EE} \approx \underline{B}\underline{K}\left(1 \!-\! \frac{\underline{\tau} \underline{K}}{\underline{S}} \right)\frac{\underline{R}_k}{{\underline{P} _{\tot}}}.
\end{equation}
This closed-form expression highlights the scalability of the proposed EE framework. The accuracy of the derived throughput approximation is validated in Fig.~\ref{fig:SE_submmW}.
Furthermore, this analytical EE formulation enables straightforward and consistent comparisons across different system configurations.

\setcounter{TempEqCnt}{\value{equation}}
\begin{figure*}[!t]
\normalsize
\begin{equation}
\underline{P}_{\tot} = \underline{\coef} _{\sfP} \underline{P} \underline{K}  + \underline{N} \sum _{i=0,1,2} \underline{\coef} _{\sfn,i} \underline{K} ^{i} + \sum _{i=0,1,3} \underline{\coef} _{\sfk,i} \underline{K} ^{i} + \underline{\coef} _{\sfR} \underline{K} \left(1 - \frac{\underline{\tau} \underline{K}}{\underline{S}} \right) {\underline{R}} + \underline{\coef} _{\fix},
    \label{eq:Ptot_sub6}
\end{equation}
\hrulefill
\end{figure*}

\subsection{Massive MIMO System on MmWave Band}
\label{sec:mmW}
In the 5G era, massive MIMO systems operating in the millimeter-wave (mmWave) frequency band have emerged as a significant milestone in the evolution of multi-antenna technologies.
Assume a massive MIMO system with bandwidth $\bar{B}$, where the BS is equipped with an array consisting of $\bar{N}$ elements and serves $\bar{K}$ single-antenna users.
Due to the sparse nature of mmWave propagation, the channel can be modeled as
\begin{equation}
    \bar{\bh} _k = \left( \sqrt{\frac{\kappa}{\kappa+1}} g _k \bb(\bar{r} _k, \bar{\varphi} _k) + \sqrt{\frac{1}{\kappa+1}} \bar{\bh} _{\NLoS}\right)\bar{\gamma} _k,
\end{equation}
where $\kappa$ is the Rician factor, $\bar{\gamma} _{k} = \bar{\lambda} / \bar{r} _k$ represents the large-scale fading coefficient, and $\bar{\bh}_{\mathrm{NLoS}}$ is modeled as a complex Gaussian vector. 

Hybrid beamforming architectures are commonly adopted in mmWave massive MIMO systems to reduce hardware cost and complexity. 
Taking the uplink transmission as an example, the received signal at the BS is expressed as
\begin{equation}
\bar{\by} _{\BS} = \sqrt{P ^{\UL}} \bar{\bW} ^{\ctrans} _{\UL} \bF ^{\ctrans} \bar{\bH} \bx _{\UE} + \bar{\bW} ^{\ctrans} _{\UL} \bF ^{\ctrans} \bn _{\BS},
\label{eq:yB_mmw}
\end{equation}
where $\bF$ denotes the analog beamforming matrix, and $\bar{\bW} ^{\UL}$ represents the digital beamforming matrix, configured similarly to \eqref{eq:Wul}, with $\bH$ replaced by the effective channel matrix $\bar{\bH}_{\eq} = \bF^{\ctrans} \bar{\bH}$.
The ergodic throughput is approximated in the following proposition.

\begin{proposition}
\label{prop:SE_mmW}
Assume that the analog beamforming matrix is configured as $\bF = \frac{1}{\sqrt{\bar{N}}}[\bb(r_{1},\varphi _1),\dots,\bb(r _K,\varphi _K)]$ and a ZF digital detector is employed. Then, the ergodic SE of an arbitrary user in a mmWave massive MIMO system is approximated by
\begin{equation}
\begin{aligned}
\bar{R} _k = \frac{\bar{C}}{(\bar{r} ^2 _{\max} - \bar{r} ^2 _{\min})\ln 2}\Bigg(\exp\left({\frac{\bar{r} ^2_{\max}}{\bar{C}}}\right) \mathrm{E}_1 \left(\frac{\bar{r} ^2_{\max}}{\bar{C}} \right)&\\
- \exp\left({\frac{\bar{r} ^2_{\min}}{\bar{C}}}\right)\mathrm{E}_1 \left(\frac{\bar{r} ^2_{\min}}{\bar{C}} \right) + \ln \left( \frac{\bar{r} ^2 _{\max}}{\bar{r} ^2 _{\min}}\right) \Bigg)&,
\end{aligned}
\label{eq:Rmmw}
\end{equation}
where $\bar{C} = \bar{P}\bar{\lambda}^2 \bar{N}/\noisevar$.
\end{proposition}
\begin{IEEEproof}
    Refer to Appendix~\ref{appendix:Rmmw}.
\end{IEEEproof} 

The total power consumption of a mmWave massive MIMO system is modeled in \eqref{eq:P_mmw}, shown at the top of page~\pageref{eq:P_mmw}.
Here, $N _{\RF}$ is the number of RF chains, and $P _{\mathsf{s}}$ denotes the power consumed by phase shifters.
Following a similar modeling approach to that of the mid-band XL-MIMO system, additional consideration is given to the hybrid beamforming architecture.
Accordingly, the EE of the system is defined as
\begin{equation}
    \overline{\EE} \approx \bar{B}\bar{K}\left(1 \!-\! \frac{\bar{\tau} \bar{K}}{\bar{S}} \right) \frac{\bar{R} _k}{ \bar{P} _{\tot} }.
\end{equation}

\setcounter{TempEqCnt}{\value{equation}}
\begin{figure*}[!t]
\normalsize
\begin{equation}
\begin{aligned}
\bar{P} _{\mathrm{sum}} = & \bar{\coef}_{\sfP}\bar{K}\bar{P}  + \bar{N} \left(
\xi^{\mathrm{ul}} \bar{P} _{\LNA} + \xi^{\DL} \bar{P} _{\PA, \mathsf{sta}} + \bar{P}_{\mathrm{syn}} + \bar{P} _{\RFckt} + \frac{8 \bar{B} \bar{K}^2 \bar{\tau}}{\bar{S} \bar{Q} _{\mathrm{B}}}\right) + \sum_{k=1}^{\bar{K}} (\bar{R} _k^{\mathrm{ul}} + \bar{R} _k^{\mathrm{dl}})  \bar{Q} _{\mathrm{DEC}}/\bar{Q} _{\BS} + \bar{P} _{\mathrm{B,fix}}\\
& + N_{\RF}\bar{N} P_{\mathsf{s}} +  N_{\RF} \left( \bar{P}_{\IFckt} \!+\! \xi^{\mathrm{ul}} \bar{P} _{\AD} \!+\! \xi^{\mathrm{dl}} \bar{P} _{\DA}  \!+\! \bar{B} \left(1 \!-\! \frac{\bar{\tau} \bar{K}}{\bar{S}} \right) \frac{8 \bar{K}}{\bar{Q}_{\mathrm{B}}} \!+\! \frac{\bar{B}}{\bar{S} \bar{Q}_{\mathrm{B}}} \left( 16 \bar{K}^2 \!+\! 2 \bar{K} \right)\right)
\!+\! \frac{8\bar{B} \bar{K}^3}{3 \bar{S} \bar{Q}_{\mathrm{B}}} \!+\! \bar{K} \bar{P}_{\UE},
\end{aligned}
\label{eq:P_mmw}
\end{equation}
\hrulefill
\end{figure*}

\begin{table}[!t]
\renewcommand{\arraystretch}{1.2}
	\caption{Parameters of Representative Multi-antenna Technologies \label{tab:simu}}
	\centering
	\begin{tabular}{m{0.9cm}<{\centering} m{1cm}<{\centering} m{1cm}<{\centering} m{1.1cm}<{\centering} m{0.4cm}<{\centering} m{0.6cm}<{\centering} m{0.8cm}<{\centering}}
	\hline
	System & $f_c$ & $N$ & $B$ & $K$ & $r_{\min}$ & $r_{\max}$ \\
	\hline
        Setup 1 & 3.5 GHz & 64 & 20 MHz & 8 & 70 m & 500 m\\
        Setup 2 & 7.5 GHz & [512, 1024, 2048] &400 MHz& [16, 32] & 70 m & 200 m\\
        Setup 3 & 28 GHz & 256 & 800 MHz & 16 & 70 m & 150 m \\
        \hline
	\end{tabular}
\end{table}

\begin{figure}[!t]
    \centering
    \includegraphics[scale=0.45]{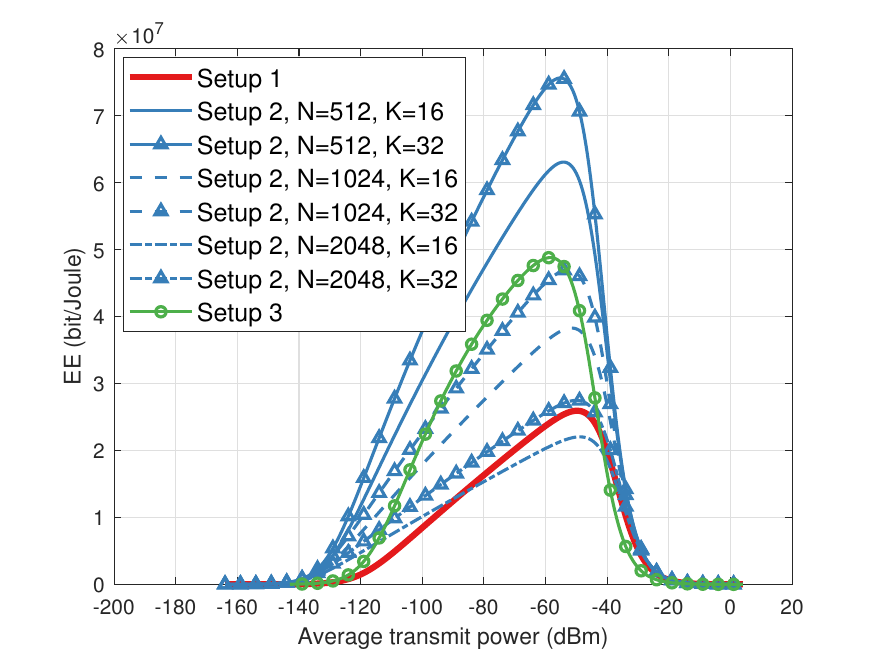}
    \caption{Comparisons of EE among representative multi-antenna technologies.}
    \label{fig:EEcmp}
\end{figure}

\subsection{Numerical Results and Discussions}

For comparisons with representative multi-antenna technologies, the simulation parameters are shown in Table~\ref{tab:simu}, where setup1, setup 2 and setup 3 simulate the mid-band XL-MIMO, Sub-6 GHz MIMO and mmW massive MIMO systems, respectively.
Fig.~\ref{fig:SE_submmW} illustrates the comparisons between the derived approximations in \eqref{eq:Rsub6} and \eqref{eq:Rmmw}, and the corresponding Monte Carlo simulations.
The results again demonstrate high accuracy of the proposed analytical expressions across different user configurations, which supports their use in EE analysis and comparisons.

The EE performance of different multi-antenna technologies under various configurations is illustrated in Fig.~\ref{fig:EEcmp}.
Hardware coefficients are set according to practical specifications.
Note that fully digital architectures are assumed for Setup 1 and Setup 2, while Setup 3 adopts a fully connected hybrid beamforming structure, with the number of RF chains satisfying $N_{\RF} = \bar{K}$.
Overall, Setup 2 achieves the highest EE among the three systems across a wide range of transmit power levels.
By contrast, Setup 1 experiences limited throughput due to its narrow bandwidth and smaller antenna array, resulting in significantly lower EE.
Although Setup 3 benefits from a large bandwidth, its EE performance is hindered by severe propagation loss and the high hardware power consumption typical of mmWave systems. 
The observed trends align with the derived EE scaling laws.
Moreover, operating with a reduced number of transceiver chains remains an energy-efficient regime, especially in channels with small angular spread. 
Regarding the small power consumption of phase shifters and RF switches relative to that of RF chains, for single-ray channels, setting $N _{\RF}\approx 2K$ (two RF chains per user) yields an EE that is near $\frac{N}{2K}$ times that of the fully digital architecture.
The EE advantage diminishes as the angular spread increases, as more RF chains are required to capture the spatial degrees of freedom.
In summary, the promising EE of mid-band XL-MIMO systems can be realized through dedicated design, particularly in the design of the numbers of antennas, users, as well as the transmit power.

\section{Conclusion}
Mid-band XL-MIMO systems, which integrate advantages in both spectral and spatial domains, are considered promising candidates for future wireless communication systems.
In this paper, an in-depth investigation of the EE of mid-band XL-MIMO systems has been presented.
A comprehensive power consumption model was first proposed to capture the impact of key system configurations.
By combining this model with closed-form approximations of the ergodic throughput, a tractable analytical framework for EE evaluation was developed.
On this basis, scaling laws of EE with respect to major system parameters were derived, providing valuable insights for energy-efficient system design.
Additionally, comparisons with representative multi-antenna technologies were conducted, which demonstrated the superiority of mid-band XL-MIMO systems under various scenarios.
Numerical results validated the accuracy of the derived scaling laws and confirmed the effectiveness of the proposed analytical framework in evaluating and guiding the EE performance of next-generation wireless systems.

{\appendices
\section{Proof of Theorem \ref{theorem:SE}}
\label{appendix:SE}

Without loss of generality, we omit the superscript $\mathsf{ul}$ and $\mathsf{dl}$. 
Applying Jensen's inequality, the throughput ${R}$ satisfies 
\begin{equation}
    {R} \leq  B \log ( 1 + \bbE\{\SINR _k\} ) \triangleq R_k ^{\ub}.
    \label{eq:Rdot}
\end{equation}
Based on the asymptotic analysis and continuous mapping theorem \cite{ConMap} and the asymptotic expression in \cite{Ali, AntSel}, we replace the exact SINR with the asymptotic expression, which derives
\begin{equation}\small
\begin{aligned}
    &\bbE\{\SINR_k\} \\
    &\overset{\mathrm{(a)}}{\leq} \bbE\left\{ \frac{P \lambda^2}{\noisevar} \left( \sum _{n\in \calN} D ^{-2} _{k,n} - (K-1)\frac{\sum _{n\in \calN} D ^{-4} _{k,n}} { \sum _{n\in \calN} D ^{-2} _{k,n} }\right) \right\}\\ \allowbreak
    & \overset{\mathrm{(b)}}{\leq} \frac{P \lambda^2}{\noisevar} \left( \bbE \left\{ \sum _{n\in \calN} D ^{-2} _{k,n} \right\} \!-\! (K\!-\!1) \frac{\bbE \left\{ \sum _{n\in \calN} D ^{-4} _{k,n} \right\}}{\bbE \left\{ \sum _{n\in \calN} D ^{-2} _{k,n} \right\}} \right),
\end{aligned}
\end{equation}
where (a) is due to the neglection of non-diagonal elements and (b) applies $\bbE\{X/Y\}\geq \bbE\{X\}/\bbE\{Y\}$.
Among them, the first item in parentheses is calculated as
\begin{equation}\small
\begin{aligned}
    \bbE & \left\{ \sum _{n\in \calN} D ^{-2} _{k,n} \right\} = \sum _{n\in \calN} \int_{r} \int _{\varphi} \frac{f(r)f(\varphi)}{r^2 - 2r \delta _n \cos \varphi + \delta _n ^2}  \rmd r \rmd \varphi\\
    &= \sum _{n\in \calN} \int _r \frac{2r}{(r^2 - \delta _n ^2) (r _{\max} ^2 - r _{\min} ^2)} \rmd r\\
    &= \sum _{n \in \calN} \frac{1}{r _{\max} ^2 - r _{\min} ^2}\ln \left( \frac{r _{\max}^2 - n^2 d_{\rmA}^2}{r _{\min} ^2 - n^2 d_{\rmA}^2} \right) \triangleq \chi.
\end{aligned}
\label{eq:Gamma}
\end{equation}
Whilst the the denominator of the second term in parentheses can be expressed as
\begin{equation}
\begin{aligned}
    \bbE \left\{ \sum _{n\in \calN} D ^{-4} _{k,n} \right\} = \sum _{n\in \calN} \bbE\{D ^{-4} _{k,n}\} \overset{(a)}{\geq} \sum _{n\in \calN} (\bbE\{D ^{-2} _{k,n}\}) ^2\\
    =\sum _{n} \left( \frac{1}{r _{\max} ^2 - r_{\min} ^2} \ln \left( \frac{r_{\max} ^2 - n^2 d_{\rmA} ^2 }{r _{\min} ^2 - n^2 d_{\rmA} ^2} \right) \right) ^2 \triangleq I,
\end{aligned}
\label{eq:I}
\end{equation}
where (a) is the application of the Jensen's inequality. 
Subsituting \eqref{eq:Gamma} and \eqref{eq:I} into \eqref{eq:Rdot}, the proof is completed.

\section{Proof of Proposition \ref{coro:SE_UB}}
\label{appendix:SE_UB}
Firstly, we simplify the summation of $n$ in $\chi$, and define
\begin{equation}
    \bar{\chi} \triangleq \frac{1}{r_{\max}^2 - r _{\min} ^2}\int _{-\frac{N}{2}} ^{\frac{N}{2}} \ln \left( \frac{r_{\max} ^2 / d _{\rmA} ^2 - n^2}{r_{\min} ^2 / d _{\rmA} ^2 - n^2} \right) \rmd n.
\end{equation}
Applying the integral property, the simplified approximation, i.e., $\chi$ in {\it Proposition \ref{coro:SE_UB}} can be obtained.
Recalling the asymptotic expression of the SINR in \cite{Ali}, we apply the following trace inequality
\begin{equation}
    \frac{\Tr(\cov_{k}\cov_j)}{\Tr(\cov _j)} \geq \frac{\min(\diag(\cov_{k})) \Tr(\cov _j)}{\Tr(\cov _j)}\geq \frac{1}{(r + \frac{Nd_{\rmA}}{2})^2}.
\end{equation}
We define $\bar{I}$ as
\begin{equation}
    \bar{I} = \int_{r_{\min}} ^{r _{\max}}\frac{1}{(r + \frac{Nd_{\rmA}}{2})^2}f(r) \rmd r.
\end{equation}
Calculating the integral as \eqref{eq:I_bar} and the simplified upper bound in \eqref{eq:Rapp1} be derived.

\section{Proof Of Proposition \ref{prop:antenna}}
\label{appendix:antenna}
When the number of antennas is small, i.e, $N\ll 2 r_{\min}/d_\rmA$, $\bar{\chi}$ can be expanded as
\begin{equation}\small
\begin{aligned}
    \bar{\chi} =& \frac{1}{(r_{\max}^2 - r_{\min} ^2)}\Bigg( 2\ln \left( \frac{r_{\max}}{ r_{\min} } \right) N + \\
    &\sum_{i=1}^{\infty}\left(\frac{1}{i} \! - \! \frac{2}{2i+1} \right)\left( \frac{d_\rmA ^{2i}}{(2 r_{\min})^{2i}} \!-\! \frac{d_\rmA ^{2i}}{(2 r_{\max})^{2i}} \right) N ^{2i+1} \Bigg)\\
    \overset{\mathrm{(a)}}{=}&\frac{2\ln \left( {r _{\max}} / { r _{\min} } \right)}{r_{\max}^2 - r_{\min} ^2} N + \mathcal{O}(N),
\end{aligned}
\label{eq:chi_app}
\end{equation}
where (a) is the conservation of the first order, due to $\frac{d ^{2i} _\rmA}{(2 r_{\min})^{2i}} \!-\! \frac{d ^{2i} _\rmA}{(2 r_{\max})^{2i}}$ is negligible.
Taking the derivative of $\bar{\chi}$ in \eqref{eq:barGamma} w.r.t. $N$, we obtain
\begin{equation}
    \frac{\rmd \bar{\chi} }{\rmd N} = \frac{1}{{r_{\max}^2 - r_{\min} ^2}}\ln \left( \frac{4r_{\max}^2 / d_{\rmA} ^2 - N ^2}{4r_{\min} ^2/d_{\rmA}^2 - N ^2} \right).
\end{equation}
It can be observed that ${\rmd \bar{\chi} }/{\rmd N} > 0$ holds and increases with $N $.
Therefore, $\bar{\chi} $ increases with $N_{\BS}$ with a steep increase.
When $N$ approaches $2r_{\min}/d_{\rmA}$, we define $N=2r_{\min}/d_{\rmA} - \varepsilon$ and submit it into $\bar{\chi}$.
Letting $\varepsilon \rightarrow 0$, the limit in \eqref{eq:chi_limit} can be derived.
Moreover, based on the approximation \eqref{eq:chi_app}, we obtain the equivalence in \eqref{eq:R_LTx} and \eqref{eq:R_HTx}.

\section{Proof of Corollary \ref{Coro:N}}
\label{appendix:N}
Regarding the limit performance of EE w.r.t. $N$, we turn to analyze $\lim _{N \rightarrow \infty} {\SINR}/{P _{\tot}}$ due to the property, $0 \leq {R} _k ^{\app} \leq \SINR$, which is further converted to calculating $\lim _{N \rightarrow \infty} { \sum _{n\in \calN} D ^{-2} _{k,n} }/{P_{\tot}}$.
Recalling the definition in \eqref{eq:Gamma} and focusing on the limit of the numerator, we have 
\begin{equation}\footnotesize
\begin{aligned}
    \lim _{N \rightarrow \infty}& \left\{ \sum _{n\in \calN} D ^{-2} _{k,n} \right\} 
    \!=\! \lim_{N\rightarrow \infty}  \sum _{n\in \calN} \int_{r} \int _{\varphi} \frac{f(r)f(\varphi)}{r^2 \!-\! 2r \delta _n \cos \varphi \!+\! \delta _n ^2}  \rmd r \rmd \varphi\\
    &\overset{(a)}{=} \int_{r} \int _{\varphi} \lim_{N\rightarrow \infty} \sum _{n\in \calN}\frac{f(r)f(\varphi)}{r^2 - 2r \delta _n \cos \varphi + \delta _n ^2}  \rmd r \rmd \varphi\\
    &\overset{(b)}{=} \int_{r} \int _{\varphi} \frac{2}{d_{\rmA}\sin \varphi (r _{\max}^2 - r _{\min}^2)} \rmd r \rmd \varphi \overset{(c)}{=} \mathrm{const},
\end{aligned}
\end{equation}
where (a) is due to Lebesgue's Dominated Convergence Theorem, (b) is the application of the near-field property in \cite{LuICC, Tianmidband}, and (c) is because $\varphi$ cannot reach $0$ and $\pi$.
Since $\lim_{N \rightarrow \infty} P _{\tot} = \infty$ is satisfied, we complete the proof.

Recalling that when the transmit power is low, the throughput and the power consumption scale with $N$ linearly.
As $N$ grows large, the EE reaches to a flat zone by the probability $\eta _{\sfn}$, satisfying
\begin{equation}
    \frac{\Delta_{\sfR} N^{\mathsf{kp}}}{\Delta_{\sfn,0} + \Delta_{\sfn,1} N^{\mathsf{kp}}} \approx \eta_{\sfn} \frac{\Delta_{\sfR} }{\Delta_{\sfn,1}},
\end{equation}
where $\Delta_{\sfR}$, $\Delta_{\sfn,0}$ and $\Delta_{\sfn,1}$ are given as
\begin{subequations}
    \begin{align}
        \Delta_{\sfR} &= \frac{BK\left(1-\frac{\tau K}{S}\right)P \lambda^2 N}{\noisevar\ln2 (r_{\max}^2 - r_{\min} ^2) } \left(1 + 2\ln\left( \frac{r_{\max}}{r _{\min}} \right) \right),\\
        \Delta_{\sfn,0} &= \coef _{\sfP}KP  + \coef_{\sfk,1} K +\coef_{\sfk,3} K ^{3} + \coef _{\fix},\\
        \Delta _{\sfn,1} &= \coef _{\sfn,0} + \coef _{\sfn,1} K + \coef _{\sfn,2} K ^{2}.
    \end{align}
\end{subequations}
Therefore, the knee point in \eqref{eq:x_kp} can be obtained.

\section{Proof of Proposition \ref{prop:Rsub6}}
\label{appendix:Rsub6}
The ergodic throughput satisfies
\begin{equation}
\begin{aligned}
    \underline{R} _k &= \bbE _{\underline{\bh} _k,\underline{\gamma} _k} \{ \underline{P} / [(\underline{\bH} ^{\ctrans} \underline{\bH}) ^{-1}] _{k,k} \}\\
    & \overset{{(a)}}{\geq} \int _{\underline{r}} \log_2\left(1 + \frac{\underline{P}}{\noisevar}(\underline{N} - \underline{K}) \underline{\gamma} \right) f(\underline{r}) \rmd \underline{r},
\end{aligned}
\end{equation}
where $(a)$ is due to the {\it Proposition 3} in \cite{EENgo}.
Calculating the integral and the lower-bound is obtained.

\section{Proof of Proposition \ref{prop:SE_mmW}}
\label{appendix:Rmmw}
When the analog beamforming and digital beamforming matrices are configured, the ergodic throughput is written as
\begin{equation}
\begin{aligned}
    \bar{R} _k & \approx \bbE _{\dot{g},\gamma} \left \{\log _2\left( 1 + \frac{\bar{P}}{\noisevar} \dot{g} \bar{\gamma} \bar{N}\right) \right\} \\
    & = \int _{\bar{r}} \int _{\dot{g}} \log _2\left( 1 + \frac{\bar{P}}{\noisevar} \dot{g} \bar{\gamma} \bar{N}\right) e^{-\dot{g}} f(\bar{r}) \rmd \dot{g} \rmd \bar{r}.
\end{aligned}
\end{equation}
Calculating the integral and the approximation is obtained.
}

\bibliographystyle{IEEEtran}
\bibliography{IEEEabrv,ref}

\end{document}